\documentclass[final,1p]{elsarticle}
\usepackage{float}
\usepackage{ifpdf}
\usepackage{graphicx}
\usepackage{dcolumn}   
\usepackage{amssymb}
\usepackage{amsmath}
\date{}
\journal{Physics Letters A}
\begin{document}



\title{On symmetry preserving and symmetry broken bright, dark and antidark soliton solutions of nonlocal nonlinear Schr\"{o}dinger equation}
\author{N. Vishnu Priya$^{1}$, M. Senthilvelan$^{2}$,  Govindan Rangarajan$^{1}$, M. Lakshmanan$^{2}$}
\address{$^1$Department of Mathematics, Indian Institute of Science, Bangalore - 560 012, Karnataka, India. 
\\$^2$Centre for Nonlinear Dynamics, School of Physics, Bharathidasan University, Tiruchirappalli - 620 024, Tamilnadu, India.\\}
\begin{abstract}
We construct symmetry preserving and symmetry broken N-bright, dark and antidark soliton solutions of a nonlocal nonlinear Schr\"{o}dinger equation.  To obtain these solutions, we use appropriate eigenfunctions in Darboux transformation (DT) method.  We present explicit one and two bright soliton solutions and show that they exhibit stable structures only when we combine the field and parity transformed complex conjugate field.  Further, we derive two dark/antidark soliton solution with the help of DT method.  Unlike the bright soliton case, dark/antidark soliton solution exhibits stable structure for the field and the parity transformed conjugate field separately.  In the dark/antidark soliton solution case we observe a contrasting behaviour between the envelope of the field and parity transformed complex conjugate envelope of the field. For a particular parametric choice, we get dark (antidark) soliton for the field while the parity transformed complex conjugate field exhibits antidark (dark) soliton.  Due to this surprising result, both the field and PT transformed complex conjugate field exhibit sixteen different combinations of collision scenario.  We classify the parametric regions of dark and antidark solitons in both the field and parity transformed complex conjugate field by carrying out relevant asymptotic analysis. Further we present $2N$-dark/antidark soliton solution formula and demonstrate that this solution may have $2^{2N}\times 2^{2N}$ combinations of collisions.
\end{abstract}
 
\maketitle


\section{Introduction} 
\label{intro}

About five years ago, Ablowitz and Musslimani have proposed the following nonlocal nonlinear Schr\"{o}dinger (NNLS) equation \cite{Ablowitz}
\begin{align}
iq_t(x,t)=q_{xx}(x,t)+2\sigma q(x,t)q^*(-x,t)q(x,t)=0,\; \sigma=\pm1,
\label{pct1}
\end{align}
where $q(x,t)$ is a slowly varying pulse envelope of the field, $x$ and $t$ represent space and time variables respectively and * denotes complex conjugation.  The NNLS equation (\ref{pct1}) is invariant under the parity-time (PT) transformation. PT symmetric systems, which allow lossless propagation due to their balance of gain and loss, have attracted considerable attention in recent years \cite{{pt1},{pt2},{pt3},{pt4},{pt5}}.  Equation (\ref{pct1}) attracted many researchers to study its physical and mathematical aspects intensively, see for example Refs. \cite{{Fokas},{AM},{zhang},{AM2},{Yan}}.  The integrability of (\ref{pct1}) is proved by (i) the existence of a Lax pair, (ii) existence of infinite number of conservation laws and (iii) existence of N-soliton solutions \cite{{Ablowitz}, {Gerdjikov}}.  The initial value problem was studied by Ablowitz et al.\cite{Ablowitz}.  Breathers, dark, antidark soliton, algebraic soliton, higher order rational solutions, periodic and hyperbolic solutions of (\ref{pct1}) have been derived for this equation in Refs. \cite{{Liming},{dad},{Ablowitz2},{chinese1},{chinese2},{Khare}}.  Discrete version of Eq. (\ref{pct1}) has also been proposed in \cite{{discrete1},{discrete},{Ma}}. Recently, Stalin and two of the present authors have constructed more general bright soliton solutions for (\ref{pct1}) by developing a nonstandard bilinearization procedure \cite{Stalin}.  In this procedure, besides Eq. (\ref{pct1}) the authors have also considered the parity transformed complex conjugate equation of (\ref{pct1}), namely
\begin{align}
iq^*_t(-x,t)=-q^*_{xx}(-x,t)-2\sigma q^*(-x,t)q(x,t)q^*(-x,t)=0,\; \sigma=\pm 1,
\label{pct2}
\end{align}
since they have assumed $q(x,t)$ and $q^*(-x,t)$ evolve independently.  Since Eq. (\ref{pct1}) is nonlocal, to evaluate the dependent variable $q(x,t)$ at $+x$, the other variable $q^*(-x,t)$ has to be evaluated at $-x$ simultaneously. The authors have obtained more general one and two soliton solutions of Eqs. (\ref{pct1}) and (\ref{pct2}) by solving them in a combined manner and studied the collision dynamics between two solitons.  The approach proposed by the authors is different from the standard one in the literature and produce a more general class of soliton solutions.  In particular, the authors have shown that the system can admit both symmetry broken solutions (the solution, $q^*(-x,t)$, which does not match with the one resulting from $q(x,t)$ after taking complex conjugation and space inversion in it) and symmetry preserving solutions (the solution, $q^*(-x,t)$, which matches with the one resulting from $q(x,t)$ after taking complex conjugation and space inversion in it).  Such broken symmetry solutions are also called spontaneously broken symmetric solutions in the literature, for example in the case of double-well parity symmetric $\phi^4$ equation. These two categories of solutions can be identified only by augmenting a separate evolution equation (\ref{pct2}) for the parity transformed complex conjugate equation in the solution process.  The nonstandard bilinearization procedure has also been applied to two coupled NNLS equations and several new localized solutions and collision dynamics have been unearthed \cite{Stalin2}.  
\par As far as the NNLS Eq. (\ref{pct1}) is concerned the symmetry broken and symmetry preserving solutions have been analyzed only for the bright soliton case.  A natural question arises in this context is what happens to the dark soliton case.  These soliton solutions for Eq. (\ref{pct1}) have already been reported in the literature \cite{{AM2},{dad}}.  However, as we pointed out above, to bring out a more general dynamical evolution of dark soliton one should consider not only Eq. (\ref{pct1}) but also Eq. (\ref{pct2}) in the solution process.  In this work, we intend to consider both the equations and construct a more general class of dark soliton solution.
\par As a by-product of this work, we also extend Darboux transformation (DT) method suitable for this class of nonlocal equations.    To make our studies a complete one, to begin with, we derive the bright soliton solution using the DT method by considering the nonlocal term $q^*(-x,t)$ as a separate quantity.  We then move on to construct dark solitons for this problem.  The dark soliton solutions which we report in this paper is a more general one and new to the literature.  In the first iteration of DT method we get two dark soliton solutions.  A careful analysis of this solution reveals that for a particular parametric choice while $q(x,t)$ exhibits dark (antidark) soliton, surprisingly, $q^*(-x,t)$ exhibits antidark (dark) soliton.  This is because, the fields $q(x,t)$ and $q^*(-x,t)$ produce dark and antidark soliton solutions in different parametric regions since they evolve independently.  In the two soliton solution, the component $q(x,t)$ has two solitons, they may have either dark or antidark soliton forms.  Therefore four types of collision between two solitons can happen in $q(x,t)$ component alone, that is (i) two dark solitons collision, (ii) antidark and dark solitons collision, (iii) dark and antidark solitons collision and (iv) two antidark solitons collision.  The $q^*(-x,t)$ component can also have these four types of collisions between the two solitons irrespective of the structures of $q(x,t)$.  Due to this novel behaviour we get $2^2\times 2^2 =16$ combinations of collision scenario in the two soliton solution alone.  This novel property can happen only by considering the symmetry broken solutions, that is by considering $q^*(-x,t)$ as a separate quantity in the solution process.  If we consider the symmetry preserving solutions they do contain only four types of collisions, because the component $q^*(-x,t)$ would have similar structure as that of $q(x,t)$.  In our studies we plot nine distinct collision structures for the two soliton solution.  By carrying out relevant asymptotic analysis of the two soliton solution we classify the parametric regions of dark and antidark solitons in both the components $q(x,t)$ and $q^*(-x,t)$.  We then derive the four soliton solution from the second iteration of the DT method. Since the solution is cumbersome we only give plots of the solution.  For the four soliton solution we get $2^4\times 2^4$ combinations of collision behaviour.  We plot some of the combinations of collision for illustration purpose.  By iterating the DT for $N$ times we get $2N$ dark soliton solution formula.  We can also generalize the collision scenario to $2N$ soliton solution and get $2^{2N}\times 2^{2N}$ combinations of collisions in it. 
  \par The plan of the paper as follows. In Sec. II, we present the DT method to construct Nth iterated solution formula for obtaining N-bright, dark and antidark soliton solutions of Eqs. (\ref{pct1}) and (\ref{pct2}).  We present explicit one and two bright soliton solutions and study the collision dynamics between two solitons in Sec. III.  In Sec. IV we construct dark and antidark soliton solutions of (\ref{pct1}) and (\ref{pct2}) and classify the parametric regions of them.  Finally we conclude our results in Sec. V.     
\section{Darboux Transformation of NNLS equation}
In this section we recall the essential ingredients of the Darboux method to construct the desired solutions.
The Lax pair of Eqs. (\ref{pct1}) and (\ref{pct2}) is given by,
\begin{align}
\Psi_x&=U\Psi  =J\Psi\Lambda +P\Psi,\nonumber\\
\Psi_t&=V\Psi =V_0\Psi\Lambda^2+V_1\Psi\Lambda+V_2\Psi,
\label{pct3}
\end{align}
where the block matrices $J$, $P$, $\Lambda$ and $\Psi$ are given by
\begin{align}
J&= \begin{pmatrix}i&0\\0&-i\end{pmatrix},\; P=\begin{pmatrix}0&iq(x,t)\\i\sigma q^*(-x,t)&0\end{pmatrix}.
\label{pct4}
\end{align}
In the above $V_0=2J$, $V_1=2P$, $V_2=JP^2-JP_x$, $\Lambda= \text{diag}(\lambda,\lambda)$, $\Psi=(\psi,\phi)^T$ and $\lambda$ is isospectral parameter.  The compatibility condition $U_t-V_x+[U,V]=0$ leads to Eqs. (\ref{pct1}) and (\ref{pct2}), where the square bracket denotes the usual commutator. 
\subsection{First Iteration of DT}
A Darboux transformation (DT) is a special gauge transformation \cite{Matveev},
\begin{eqnarray}            
\Psi[1]=T[1]\Psi = \Psi\Lambda-S_1\Psi,
\label{pct5}
\end{eqnarray}
where $\Psi$ and $\Psi[1]$ are old and new eigenfunctions of (\ref{pct3}), $T[1]$ is the DT matrix and $S_1$ is a non-singular $2\times 2$ matrix.  The DT (\ref{pct5}) transforms the original Lax pair (\ref{pct3}) into a new Lax pair,
\begin{eqnarray}  
\Psi[1]_x&=U[1]\Psi[1]&=J\Psi[1]\Lambda +P[1]\Psi[1],\nonumber\\
\Psi[1]_t&=V[1]\Psi[1]&=V_0[1]\Psi[1]\Lambda^2+V_1[1]\Psi[1]\Lambda+V_2[1]\Psi[1],
\label{pct6}
\end{eqnarray}
in which the matrices $P[1]$, $V_0[1]$, $V_1[1]$ and $V_2[1]$ assume the same forms as that of $P$, $V_0$, $V_1$ and $V_2$ except that the potentials $q(x,t)$ and $q^*(-x,t)$ have now acquired new expressions, namely $q[1](x,t)$ and $q[1]^*(-x,t)$ in $U[1]$ and $V[1]$.  Substituting the transformation (\ref{pct5}) into (\ref{pct6}) and comparing the resultant expressions with (\ref{pct3}), we find
\begin{eqnarray}            
U[1]=(T[1]_x+T[1]U)T[1]^{-1},\;\; V[1]=(T[1]_t+T[1]V)T[1]^{-1}.
\label{pct7}
\end{eqnarray}
Plugging the expressions $U[1]$, $V[1]$, $U$, $V$ and $T[1]$ in Eq. (\ref{pct6}) and equating the coefficients of various powers of $\Lambda$ on both sides, we get the following relations between old and new potentials, namely
\begin{subequations}
\begin{align}
V_0[1]&= V_0, \label{pct10} \\
V_1[1]&= V_1+[V_0,S_1],\label{pct11} \\
V_2[1]&= V_2+[V_1,S_1]+[V_0,S_1]S_1,\label{pct12} \\
P[1] & = P+[J,S_1],\label{pct13} \\
S_{1x} & = [P,S_1]+[J,S_1]S_1, \label{pct14} \\
S_{1t} & = [V_2,S_1]+[V_1,S_1]S_1+[V_0,S_1]S_1^2.
\label{pct15}
\end{align}
\label{pc15}
\end{subequations}
\par The eigenvalue problem given in (\ref{pct3}) remains invariant under the transformation (\ref{pct5}) provided $S_1$ satisfies all the Eqs. (\ref{pct10})-(\ref{pct15}).
We assume a general form for the matrix $S_1$, namely
\begin{align}
 S_1=\begin{pmatrix}S_{11}&S_{12}\\S_{21}&S_{22}\end{pmatrix}.
\label{pctS}
\end{align}
Substituting the assumed form of $S_1$ in Eq. (\ref{pct13}) and equating the matrix elements on both sides, we find
\begin{eqnarray}
q[1](x,t)=q(x,t)+2S_{12}, \;\; q[1]^*(-x,t)=q^*(-x,t)-2\sigma S_{21}.
\label{p1q1}
\end{eqnarray}
\par To obtain two parameter family of symmetry preserving and symmetry broken solutions of NNLS equations (\ref{pct1}) and (\ref{pct2}) we consider $S_1$ to be
\begin{align}
S_1=\Psi_1\Lambda_1 \Psi_1^{-1},
\label{pct18}
\end{align}
where $\Psi_1$ is the solution of (\ref{pct3}) at $\Lambda=\Lambda_1$.  The exact forms of $\Psi_1$ and $\Lambda_1$ are given by,
\begin{align}
\Psi_1=\begin{pmatrix}\psi_1(x,t)&\psi_1^*(-x,t)\\\phi_1(x,t)&\phi_1^*(-x,t)\end{pmatrix},\;\Lambda_1=\begin{pmatrix}\lambda_1&0\\0&\overline{\lambda_1}\end{pmatrix},
\label{pct19}
\end{align}
where $(\psi_1(x,t),\phi_1(x,t))^T$ is the solution of (\ref{pct3}) at $\lambda=\lambda_1$. Since we consider $q^*(-x,t)$ as a separate quantity we assume $(\psi_1^*(-x,t),\phi_1^*(-x,t))^T$ is the appropriate solution of (\ref{pct3}) at $\lambda=\overline{\lambda_1}$, where $\overline{\lambda_1}$ is an isospectral parameter.
\par Next we shall prove that the above matrix $S_1$ satisfies expressions (\ref{pct10})-(\ref{pct12}) together with (\ref{pct14}) and (\ref{pct15}).  If $\Psi_1$ is solution of eigenvalue equations (\ref{pct3}) then one can write them as
\begin{align}
\Psi_{1x}&  =J\Psi_1\Lambda_1 +P\Psi_1,\nonumber\\
\Psi_{1t}& =V_0\Psi_1\Lambda_1^2+V_1\Psi_1\Lambda_1+V_2\Psi_1.
\label{pctpsi1}
\end{align}
By considering the form of $S_1$ as in Eq. (\ref{pct18}) and rewriting the above Eqs. (\ref{pctpsi1}), we get
\begin{align}
S_{1x}&=[\Psi_{1x}\Psi_1^{-1},S_1]=[JS_1+P,S_1],\nonumber\\
S_{1t}&=[\Psi_{1t}\Psi_1^{-1},S_1]=[V_2,S_1]+[V_1,S_1]S_1+[V_0,S_1]S_1^2.
\end{align}
The above equations exactly match with the equations given in (\ref{pct14}) and (\ref{pct15}).  Using the relation (\ref{pct13}), together with the expressions given in (\ref{pct18}), the Eqs. (\ref{pct10}) - (\ref{pct12}) are all satisfied.  Thus the DT (\ref{pct5}) preserves the forms of the Lax pair associated with the NNLS equation (\ref{pct1}) and (\ref{pct2}).  Equation (\ref{pct12}) establishes the relationship between new and original potentials. 
\par The first iterated DT is given by $\Psi[1]=T[1]\Psi=\Psi\Lambda-S[1]\Psi$ (vide Eq.(\ref{pct5})).  If $\Psi_1(x,t)$ is the solution of $\Psi$ at $\Lambda=\Lambda_1$ then it should satisfy 
\begin{align}
\Psi_1[1](x,t)=T[1]\Psi_1(x,t)=0 \Rightarrow S_1 \Psi_1(x,t)=\Psi_1(x,t)\Lambda_1.
\label{pct21}
\end{align}
Expressing Eq. (\ref{pct21}) in matrix form and using Cramer's rule we can determine the exact forms of $S_{12}$ and $S_{21}$ which are given by
\begin{align}
S_{12}=\frac{(\overline{\lambda_1}-\lambda_1)\psi_1(x,t)\psi_1^*(-x,t)}{\psi_1(x,t)\phi_1^*(-x,t)-\phi_1(x,t)\psi_1^*(-x,t)},\;S_{21}=\frac{(\lambda_1-\overline{\lambda_1})\phi_1(x,t)\phi_1^*(-x,t)}{\psi_1(x,t)\phi_1^*(-x,t)-\phi_1(x,t)\psi_1^*(-x,t)}.
\label{pct24}
\end{align}
\par From (\ref{pct24}) it is evident that to determine $S_{12}$ and $S_{21}$ one should know the explicit expressions of $\psi_1(x,t)$, $\phi_1(x,t)$, $\psi_1^*(-x,t)$ and $\phi_1^*(-x,t)$ which are the solutions of the eigenvalue problem (\ref{pct3}).
Solving (\ref{pct3}) with appropriate seed solution $q(x,t)$ and $q^*(-x,t)$, one can obtain the explicit expressions of $\psi_1(x,t)$, $\phi_1(x,t)$, $\psi_1^*(-x,t)$ and $\phi_1^*(-x,t)$. With the known expressions of $\psi_1(x,t)$, $\phi_1(x,t)$, $\psi_1^*(-x,t)$ and $\phi_1^*(-x,t)$ the matrix elements $S_{12}$ and $S_{21}$ can now be fixed. Plugging the latter into (\ref{p1q1}), we obtain the formula for two parameter symmetry preserving and symmetry broken solutions for Eqs. (\ref{pct1}) and (\ref{pct2}) in the form
\begin{eqnarray}
q[1](x,t)&=&q(x,t)+2\frac{(\overline{\lambda_1}-\lambda_1)\psi_1(x,t)\psi_1^*(-x,t)}{\psi_1(x,t)\phi_1^*(-x,t)-\phi_1(x,t)\psi_1^*(-x,t)}, \nonumber\\
q[1]^*(-x,t)&=&q^*(-x,t)-2\sigma\frac{(\lambda_1-\overline{\lambda_1})\phi_1(x,t)\phi_1^*(-x,t)}{\psi_1(x,t)\phi_1^*(-x,t)-\phi_1(x,t)\psi_1^*(-x,t)}.
\label{p1q1new}
\end{eqnarray}
Through the formula (\ref{p1q1new}) we can generate symmetry preserving and symmetry broken one bright and two dark/antidark soliton solutions of (\ref{pct1}) and (\ref{pct2}). 

\subsection{Second Iteration of DT}
Second iteration of DT can be written as \cite{Matveev}
\begin{align}
\Psi[2]=\Psi\Lambda^2+\sigma_1\Lambda\Psi+\sigma_2\Psi,
\label{2dt1}
\end{align}
where $\sigma_1=-(S_1+S_2)$, $\sigma_2=S_1S_2$, $S_j=\Lambda-\Psi_j\Lambda_j\Psi_j^{-1}$, $j=1,2.$
If $\Psi_j$, $j=1,2,$ is solution of $\Psi[2]$ at $\Lambda=\Lambda_j$ then it should satisfy
\begin{align}
\Psi_j[2]=0\; \Rightarrow \Psi_j\Lambda_j^2+\sigma_1\Lambda_j\Psi_j+\sigma_2\Psi_j=0,
\label{2dt2}
\end{align}
where $\Psi_j$ and $\Lambda_j$ are given by
\begin{align}
\Psi_j=\begin{pmatrix}\psi_j(x,t)&\psi_j^*(-x,t)\\\phi_j(x,t)&\phi_j^*(-x,t)\end{pmatrix},\;\Lambda_j=\begin{pmatrix}\lambda_j&0\\0&\overline{\lambda_j}\end{pmatrix},\;j=1,2.
\label{2dt3}
\end{align}
The second iteration of DT provides us a new solution in the form
\begin{align}
q[2](x,t)=q(x,t)-2[\sigma_1]_{12},\;\;q[2]^*(-x,t)=q^*(-x,t)+2[\sigma_1]_{21}.
\label{secondit}
\end{align}
Expressing Eq. (\ref{2dt2}) in matrix elements and using Cramer's rule we can find the exact forms of $[\sigma_1]_{12}$ and $[\sigma_1]_{21}$
as
\begin{eqnarray}
\sigma_{12}&= \frac{\left|\begin{array}{cccc}\psi_1(x,t)\lambda_1^2&\psi_2(x,t)\lambda_2^2&\psi_1^*(-x,t)\overline{\lambda_1}^2&\psi_2^*(-x,t)\overline{\lambda_2}^2\\\psi_1(x,t)\lambda_1&\psi_2(x,t)\lambda_2&\psi_1^*(-x,t)\overline{\lambda_1}&\psi_2^*(-x,t)\overline{\lambda_2}
\\\psi_1(x,t)&\psi_2(x,t)&\psi_1^*(-x,t)&\psi_2^*(-x,t)\\\phi_1(x,t)&\phi_2(x,t)&\phi_1^*(-x,t)&\phi_2^*(-x,t)\end{array}\right|}{\left|\begin{array}{cccc}\psi_1(x,t)\lambda_1&\psi_2(x,t)\lambda_2&\psi_1^*(-x,t)\overline{\lambda_1}&\psi_2^*(-x,t)\overline{\lambda_2}
\\\psi_1(x,t)&\psi_2(x,t)&\psi_1^*(-x,t)&\psi_2^*(-x,t)\\\phi_1(x,t)\lambda_1&\phi_2(x,t)\lambda_2&\phi_1^*(-x,t)\overline{\lambda_1}&\phi_2^*(-x,t)\overline{\lambda_2}\\\phi_1(x,t)&\phi_2(x,t)&\phi_1^*(-x,t)&\phi_2^*(-x,t)\end{array}\right|},\nonumber\\
\sigma_{21}&= \frac{\left|\begin{array}{cccc}\phi_1(x,t)\lambda_1^2&\phi_2(x,t)\lambda_2^2&\phi_1^*(-x,t)\overline{\lambda_1}^2&\phi_2^*(-x,t)\overline{\lambda_2}^2\\\phi_1(x,t)\lambda_1&\phi_2(x,t)\lambda_2&\phi_1^*(-x,t)\overline{\lambda_1}&\phi_2^*(-x,t)\overline{\lambda_2}
\\\phi_1(x,t)&\phi_2(x,t)&\phi_1^*(-x,t)&\phi_2^*(-x,t)\\\psi_1(x,t)&\psi_2(x,t)&\psi_1^*(-x,t)&\psi_2^*(-x,t)\end{array}\right|}{\left|\begin{array}{cccc}\psi_1(x,t)\lambda_1&\psi_2(x,t)\lambda_2&\psi_1^*(-x,t)\overline{\lambda_1}&\psi_2^*(-x,t)\overline{\lambda_2}
\\\psi_1(x,t)&\psi_2(x,t)&\psi_1^*(-x,t)&\psi_2^*(-x,t)\\\phi_1(x,t)\lambda_1&\phi_2(x,t)\lambda_2&\phi_1^*(-x,t)\overline{\lambda_1}&\phi_2^*(-x,t)\overline{\lambda_2}\\\phi_1(x,t)&\phi_2(x,t)&\phi_1^*(-x,t)&\phi_2^*(-x,t)\end{array}\right|}.
\label{fidet}
\end{eqnarray}
Substituting (\ref{fidet}) in (\ref{secondit}) we can get the second iterated DT solution formula. Using this formula we can obtain two bright and dark soliton solutions of (\ref{pct1}) and (\ref{pct2}).    
\subsection{$N$th Iteration of DT}
$N$th iteration of DT can be written as 
\begin{align}
\Psi[N]=T[N]\Psi=\Psi\Lambda^N+\sigma_1[N]\Lambda^{N-1}\Psi+\sigma_2[N]\Lambda^{N-2}+\cdots+\sigma_N[N]\Psi.
\label{ndt1}
\end{align}
If $\Psi_j$, $j=1,2,\cdots,N,$ is solution of $\Psi[j]$ at $\Lambda=\Lambda_j$, it should satisfy
\begin{align}
\Psi_j[N]=0\; \Rightarrow \Psi\Lambda^N+\sigma_1[N]\Lambda^{N-1}\Psi+\sigma_2[N]\Lambda^{N-2}+\cdots+\sigma_N[N]\Psi=0.
\label{ndt2}
\end{align}
In the above $\sigma_1[N]=-(S_1+S_2+\cdots+S_N)$, $\sigma_2[N]=S_1S_2+S_1S_3+S_2S_3+\cdots+S_{N-1}S_N$, $\sigma_N[N]=S_1S_2\cdots S_{N-1}S_N$, $S_j=\Lambda-\Psi_j\Lambda_j\Psi_j^{-1}$, $j=1,2,\cdots,N$ and $\Psi_j$, $\Lambda_j$ are given by
\begin{align}
\Psi_j=\begin{pmatrix}\psi_j&\overline{\psi_j}\\\phi_j&\overline{\phi_j}\end{pmatrix},\;\Lambda_j=\begin{pmatrix}\lambda_j&0\\0&\overline{\lambda_j}\end{pmatrix}, \;\;j=1,2,\cdots,N.
\label{ndt3}
\end{align}
Nth iteration of DT leads us to a new solution of the form
\begin{align}
q[N](x,t)=q(x,t)-2[\sigma_1[N]]_{12},\;\;q[N]^*(-x,t)=q^*(-x,t)+2[\sigma_1[N]]_{21}.
\label{nit}
\end{align}
Expressing Eq. (\ref{ndt2}) in matrix elements and using Cramer's rule we can find exact forms of $[\sigma_1[N]]_{12}$ and $[\sigma_1[N]]_{21}$
as
\begin{eqnarray}
[\sigma_1[N]]_{12}= \frac{|\Delta_2|}{|\Delta_1|},\;\;[\sigma_1[N]]_{21}= \frac{|\Delta_3|}{|\Delta_1|},
\label{ndt4}
\end{eqnarray}
where $\Delta_1$, $\Delta_2$ and $\Delta_3$ are given by
\begin{eqnarray}
\Delta_1 = \scriptsize\left|\begin{array}{cccccc}
\psi_1(x,t)\lambda_1^{N-1}&\cdots&\psi_N(x,t)\lambda_N^{N-1}&\psi_1^*(-x,t)\overline{\lambda_1}^{N-1}&\cdots & \psi_N^*(-x,t)\overline{\lambda}_N^{N-1}\\
\cdots&\cdots&\cdots&\cdots&\cdots&\cdots \\\psi_1(x,t)&\cdots&\psi_N(x,t)&\psi_1^*(-x,t)&\cdots & \psi_N^*(-x,t)\\\phi_1(x,t)\lambda_1^{N-1}&\cdots&\phi_N(x,t)\lambda_N^{N-1}&\phi_1^*(-x,t)\overline{\lambda_1}^{N-1}&\cdots & \phi_N^*(-x,t)\overline{\lambda}_N^{N-1}\\
\cdots&\cdots&\cdots&\cdots&\cdots\\\phi_1(x,t)&\cdots & \phi_N(x,t)&\phi_1^*(-x,t)&\cdots&\phi_N^*(-x,t)\end{array} \right|,
\end{eqnarray}
\begin{eqnarray}
\Delta_2 = \scriptsize\left|\begin{array}{cccccc}
\psi_1(x,t)\lambda_1^{N}&\cdots&\psi_N(x,t)\lambda_N^{N}&\psi_1^*(-x,t)\overline{\lambda_1}^{N}&\cdots & \psi_N^*(-x,t)\overline{\lambda}_N^{N}\\
\cdots&\cdots&\cdots&\cdots&\cdots&\cdots \\\psi_1(x,t)&\cdots&\psi_N(x,t)&\psi_1^*(-x,t)&\cdots & \psi_N^*(-x,t)\\\phi_1(x,t)\lambda_1^{N-2}&\cdots&\phi_N(x,t)\lambda_N^{N-2}&\phi_1^*(-x,t)\overline{\lambda_1}^{N-2}&\cdots & \phi_N^*(-x,t)\overline{\lambda}_N^{N-2}\\
\cdots&\cdots&\cdots&\cdots&\cdots\\\phi_1(x,t)&\cdots & \phi_N(x,t)&\phi_1^*(-x,t)&\cdots&\phi_N^*(-x,t)\end{array} \right|,
\end{eqnarray}
\begin{eqnarray}
\Delta_3 = \scriptsize\left|\begin{array}{cccccc}
\phi_1(x,t)\lambda_1^{N}&\cdots&\phi_N(x,t)\lambda_N^{N}&\phi_1^*(-x,t)\overline{\lambda_1}^{N}&\cdots & \phi_N^*(-x,t)\overline{\lambda}_N^{N}\\
\cdots&\cdots&\cdots&\cdots&\cdots&\cdots \\\phi_1(x,t)&\cdots&\phi_N(x,t)&\phi_1^*(-x,t)&\cdots & \phi_N^*(-x,t)\\\psi_1(x,t)\lambda_1^{N-2}&\cdots&\psi_N(x,t)\lambda_N^{N-2}&\psi_1^*(-x,t)\overline{\lambda_1}^{N-2}&\cdots & \psi_N^*(-x,t)\overline{\lambda}_N^{N-2}\\
\cdots&\cdots&\cdots&\cdots&\cdots\\\psi_1(x,t)&\cdots & \psi_N(x,t)&\psi_1^*(-x,t)&\cdots&\psi_N^*(-x,t)\end{array} \right|.
\end{eqnarray}
Substituting (\ref{ndt4}) in (\ref{nit}) one can get $N$th DT solution formula. Using this formula we can obtain symmetry preserving and symmetry broken $N$-bright and $2N$-dark soliton solutions of (\ref{pct1}) and (\ref{pct2}). 
\section{Bright soliton solutions of NNLS equation}
\subsection{One bright soliton solution}
In this subsection, we construct the one bright soliton solution of Eqs. (\ref{pct1}) and (\ref{pct2}). To construct them, we feed a vacuum solution, that is $q(x,t)=q^*(-x,t)=0$ as seed solution to the focusing NNLS equation ($\sigma=+1$). 
By solving the Lax pair equations (\ref{pct3}) with $q(x,t)=q^*(-x,t)=0$ and by considering the forms of $\Psi_1$ and $\Lambda_1$ as given in Eq. (\ref{pct19}) one can get the symmetry broken eigenfunctions of the form $\psi_1(x,t)=c_{11}e^{\eta_1}$, $\psi_1^*(-x,t)=\overline{c_{11}}e^{\overline{\eta_1}}$, $\phi_1(x,t)=c_{21}e^{-\eta_1}$ and $\phi_1^*(-x,t)=\overline{c_{21}}e^{\overline{-\eta_1}}$, with $\eta_1=i\lambda_1x+2i\lambda_1^2t$ and $\overline{\eta_1}=i\overline{\lambda_1}x+2i\overline{\lambda_1}^2t$.  Substituting these basic solutions in the first iterated DT formula (\ref{p1q1new}) one can generate two parameter family of symmetry broken one bright soliton solution of (\ref{pct1}) and (\ref{pct2}) as
\begin{align}
q[1](x,t)=&\frac{2(\overline{\lambda_1}-\lambda_1)\frac{c_{11}}{c_{21}}e^{2\eta_1}}{e^{2(\eta_1-\overline{\eta_1})+R}-1},\nonumber\\
q[1]^*(-x,t)=&\frac{2(\overline{\lambda_1}-\lambda_1)\frac{\overline{c_{21}}}{\overline{c_{11}}}e^{-2\overline{\eta_1}}}{e^{2(\eta_1-\overline{\eta_1})+R}-1},
\label{onesol}
\end{align}
where $R=\ln\left(\frac{c_{11}\overline{c_{21}}}{c_{21}\overline{c_{11}}}\right)$. The solution (\ref{onesol}) is the more general one soliton solution of (\ref{pct1}) and (\ref{pct2}).  We call it as symmetry broken solution (except for specific parameter values given below) since $q[1](x,t)$ and $q[1]^*(-x,t)$ are independent and cannot deduce one from the other.  By choosing $\lambda_1=\frac{\overline{k}_1}{2}$, $\overline{\lambda_1}=\frac{-k_1}{2}$, $c_{11}=\alpha e^{\bar{\xi^{(0)}}}$, $\overline{c_{11}}=(k_1+\overline{k}_1)$, $c_{21}=(k_1+\overline{k}_1)$ and $\overline{c_{21}}=\beta e^{\xi^{(0)}}$  in (\ref{onesol}), we can get the solution which is presented in \cite{Stalin}.  One can also deduce symmetry preserving one soliton solution of NNLS equation from (\ref{onesol}) by confining the parametric conditions in the form $\overline{\lambda_1}=-i\delta_1$, $\lambda_1=i\overline{\delta}_1$, $c_{11}=ie^{i\bar{\theta}_1}$, $\overline{c_{21}}=ie^{i\theta_1}$, $c_{21}=1$ and $\overline{c_{11}}=1$, where $\delta_1$, $\overline{\delta}_1$, $\theta_1$ and $\bar{\theta}_1$ are real constants.  As a result, we obtain
\begin{align}
q[1](x,t)=-\frac{2(\delta_1+\overline{\delta}_1)e^{i\bar{\theta}_1}e^{-2\overline{\delta}_1x-4i\overline{\delta}_1^2t}}{e^{i(\theta_1+\bar{\theta}_1)}e^{-2(\delta_1+\overline{\delta}_1)x-4i(\overline{\delta}_1^2-\delta_1^2)t}+1}.
\label{ablo}
\end{align}
The above solution coincides with the one given in \cite{Ablowitz}.  One can get $q[1]^*(-x,t)$ from (\ref{ablo}) by taking complex conjugate of it and inversing the space variable in it.
The symmetry broken solution (\ref{onesol}) for generic parametric conditions can also be written in terms of trigonometric functions as
\begin{align}
q[1](x,t)=&\frac{(\overline{\lambda_1}-\lambda_1)\frac{c_{11}}{c_{21}}e^{\eta_1+\overline{\eta_1}-\frac{R}{2}}}{i\sin[\eta_1-\overline{\eta_1}-\frac{iR}{2}]},\nonumber\\
q[1]^*(-x,t)=&\frac{2(\overline{\lambda_1}-\lambda_1)\frac{\overline{c_{21}}}{\overline{c_{11}}}e^{-\eta_1-\overline{\eta_1}-\frac{R}{2}}}{i\sin[\eta_1-\overline{\eta_1}-\frac{iR}{2}]}.
\label{onesol1}
\end{align}
\begin{figure}
\includegraphics[width=\linewidth]{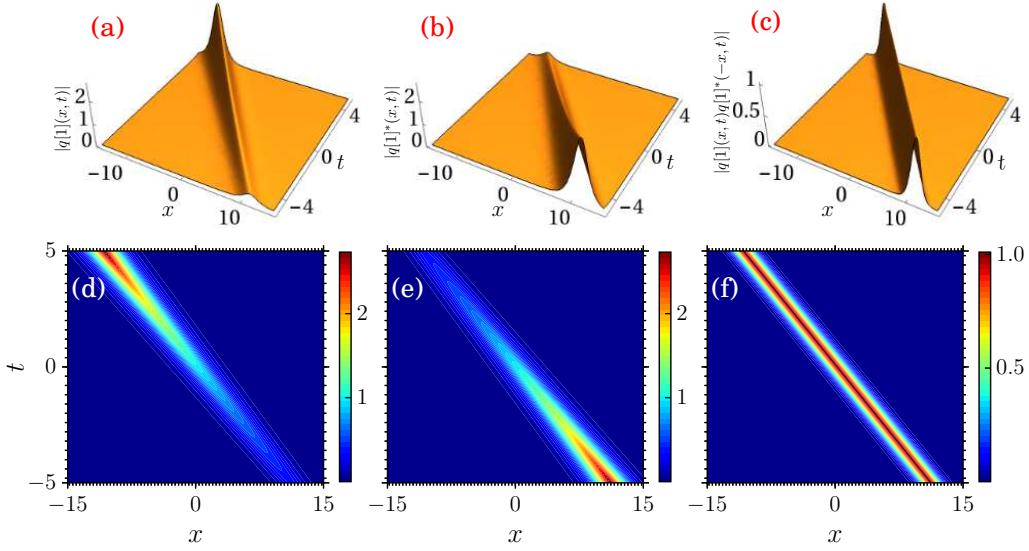}
\caption{One bright soliton evolution of NNLS system (The values of parameters are given in text).}
\end{figure}
\par The above solution (\ref{onesol1}) develops singularity at $x=0$, $t=\frac{n\pi+i\frac{R}{2}}{2(\lambda_1^2-\overline{\lambda_1}^2)}$.  This one bright soliton solution is plotted in Fig. 1 for the parameter values $\lambda_1=0.6-0.5i$, $\overline{\lambda_1}=0.5+0.5i$, $c_{11}=0.5+i$, $c_{21}=-1-i$, $\overline{c_{11}}=1-i$, $\overline{c_{21}}=0.5-i$. The absolute value of $q_1(x,t)$ plotted in Fig. 1(a) shows that the amplitude of the soliton decays at $t\to\infty$ in the $+x$ direction.  In contrast, the absolute value of parity transformed conjugate field grows at $t\to\infty$ in the $+x$ direction and is illustrated in the Fig.1(b).  We can get a stable propagation of soliton for the absolute value of $|q(x,t)q^*(-x,t)|$ which is demonstrated in Fig.1(c).  

\subsection{Two bright soliton solution}
Now we derive the two bright soliton solution of Eqs. (\ref{pct1}) and (\ref{pct2}).  For this, we have to solve the Lax pair equations (\ref{pct3}) at $\Lambda=\Lambda_j$, $j=1,2,$ with the seed solutions $q(x,t)=q^*(-x,t)=0$.  By doing so, we obtain the basic solutions of the form $\psi_j(x,t)=c_{1j}e^{\eta_j}$, $\psi_j^*(-x,t)=\overline{c_{1j}}e^{\overline{\eta_j}}$, $\phi_j(x,t)=c_{2j}e^{-\eta_j}$ and $\phi_j^*(-x,t)=\overline{c_{2j}}e^{\overline{-\eta_j}}$, where $\eta_j=i\lambda_jx+2i\lambda_j^2t$ and $\overline{\eta_j}=i\overline{\lambda_j}x+2i\overline{\lambda_j}^2t$, $j=1,2$, where $c_{ij}$ and $\overline{c_{ij}}$, $i=1,2$ are integration constants.  Substituting these solutions into the second iterated DT formula (\ref{secondit}) and simplifying the resultant expressions, we can obtain two bright soliton solution which is of the form
\begin{align}
q[2](x,t)=&\frac{-2}{D_1}\bigg(c_{11}\overline{c_{11}}c_{12}\overline{c_{22}}(\lambda_1-\overline{\lambda_1})(\lambda_1-\lambda_2)(\overline{\lambda_1}-\lambda_2)e^{\eta_1+\overline{\eta_1}+\eta_2-\overline{\eta_2}}\nonumber\\
&-c_{11}\overline{c_{11}}c_{22}\overline{c_{12}}(\lambda_1-\overline{\lambda_1})(\lambda_1-\overline{\lambda_2})(\overline{\lambda_1}-\overline{\lambda_2})e^{\eta_1+\overline{\eta_1}-\eta_2+\overline{\eta_2}}\nonumber\\
&-c_{12}\overline{c_{11}}c_{21}\overline{c_{12}}(\overline{\lambda_1}-\lambda_2)(\overline{\lambda_1}-\overline{\lambda_2})(\lambda_2-\overline{\lambda_2})e^{-\eta_1+\overline{\eta_1}+\eta_2+\overline{\eta_2}}\nonumber\\
&+c_{11}\overline{c_{12}}c_{12}\overline{c_{21}}(\lambda_1-\lambda_2)(\lambda_1-\overline{\lambda_2})(\lambda_2-\overline{\lambda_2})
e^{\eta_1-\overline{\eta_1}+\eta_2+\overline{\eta_2}}\bigg),\nonumber\\
q[2]^*(-x,t)=&\frac{2}{D_1}\bigg(c_{21}\overline{c_{12}}c_{22}\overline{c_{21}}(\lambda_1-\overline{\lambda_1})(\lambda_1-\lambda_2)(\overline{\lambda_1}-\lambda_2)e^{-\eta_1-\overline{\eta_1}-\eta_2+\overline{\eta_2}}\nonumber\\
&-c_{12}\overline{c_{21}}c_{21}\overline{c_{22}}(\lambda_1-\overline{\lambda_1})(\lambda_1-\overline{\lambda_2})(\overline{\lambda_1}-\overline{\lambda_2})e^{-\eta_1-\overline{\eta_1}+\eta_2-\overline{\eta_2}}\nonumber\\
&-c_{11}\overline{c_{21}}c_{22}\overline{c_{22}}(\overline{\lambda_1}-\lambda_2)(\overline{\lambda_1}-\overline{\lambda_2})(\lambda_2-\overline{\lambda_2})e^{\eta_1-\overline{\eta_1}-\eta_2-\overline{\eta_2}}\nonumber\\
&+c_{21}\overline{c_{11}}c_{22}\overline{c_{22}}(\lambda_1-\lambda_2)(\lambda_1-\overline{\lambda_2})(\lambda_2-\overline{\lambda_2})
e^{-\eta_1+\overline{\eta_1}-\eta_2-\overline{\eta_2}}\bigg),\nonumber
\end{align}
\begin{align}
D_1=&-c_{12}\overline{c_{12}}c_{21}\overline{c_{21}}(\lambda_1-\overline{\lambda_1})(\lambda_2-\overline{\lambda_2})
e^{-\eta_1-\overline{\eta_1}+\eta_2+\overline{\eta_2}}
\nonumber\\
&-c_{11}\overline{c_{11}}c_{22}\overline{c_{22}}(\lambda_1-\overline{\lambda_1})(\lambda_2-\overline{\lambda_2})
e^{\eta_1+\overline{\eta_1}-\eta_2-\overline{\eta_2}}\nonumber\\
&+c_{21}\overline{c_{11}}c_{22}\overline{c_{12}}(\lambda_1-\lambda_2)(\overline{\lambda_1}-\overline{\lambda_2})
e^{-\eta_1+\overline{\eta_1}-\eta_2+\overline{\eta_2}}\nonumber\\
&-c_{11}\overline{c_{12}}c_{22}\overline{c_{21}}(\overline{\lambda_1}-\lambda_2)(\lambda_1-\overline{\lambda_2})
e^{\eta_1-\overline{\eta_1}-\eta_2+\overline{\eta_2}}\nonumber\\
&-c_{12}\overline{c_{11}}c_{21}\overline{c_{22}}(\overline{\lambda_1}-\lambda_2)(\lambda_1-\overline{\lambda_2})
e^{-\eta_1+\overline{\eta_1}+\eta_2-\overline{\eta_2}}\nonumber\\
&+c_{11}\overline{c_{21}}c_{12}\overline{c_{22}}(\lambda_1-\lambda_2)(\overline{\lambda_1}-\overline{\lambda_2})
e^{\eta_1-\overline{\eta_1}+\eta_2-\overline{\eta_2}}.
\label{2sol}
\end{align}
\begin{figure}
\includegraphics[width=\linewidth]{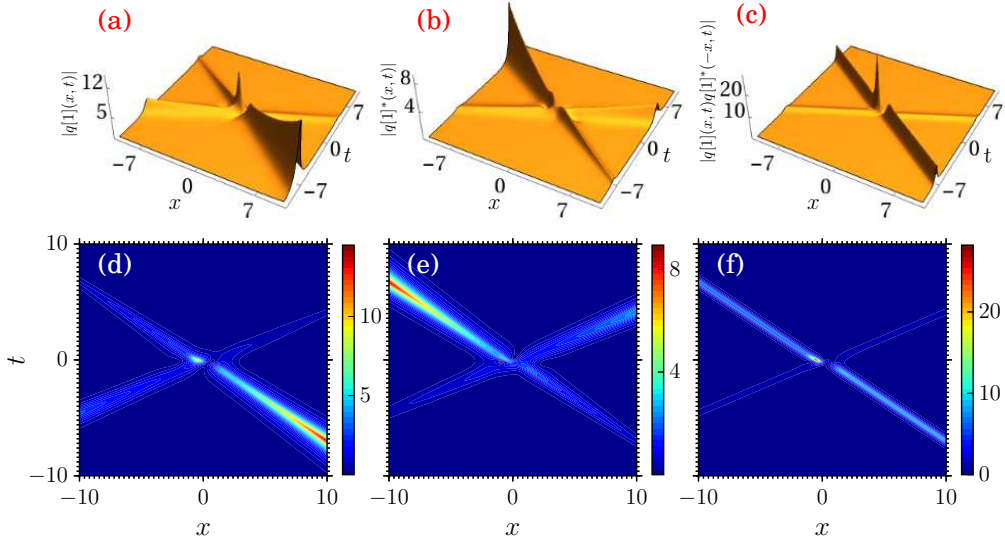}
\caption{Bright two soliton evolution of NNLS system for the parameter values $\lambda_1=-0.5+0.5i$, $\overline{\lambda_1}=-0.6-0.5i$, $\lambda_2=0.4+0.5i$, $\overline{\lambda_2}=0.3-0.5i$, $c_{11}=0.5+0.25i$, $\overline{c_{11}}=-0.25+0.25i$, $c_{21}=0.25+0.25i$, $\overline{c_{21}}=0.5-0.25i$, $c_{12}=0.25+0.25i$, $\overline{c_{12}}=-0.25-0.25i$, $c_{22}=0.25+0.25i$, $\overline{c_{22}}=0.25-0.25i$.}
\end{figure}
\par The two bright soliton solution (\ref{2sol}) of NNLS equation is plotted in Fig.2.  As in the one soliton case, two soliton also exhibits stable propagation only for the absolute value of $|q(x,t)q^*(-x,t)|$, which can be clearly seen in Figs. 2.  This more general two soliton solution (\ref{2sol}) also coincides with the factorized form of two soliton solution given in \cite{Stalin}.     
\par If we consider $2N$ set of basic solutions $\psi_j(x,t)=c_{1j}e^{\eta_j}$, $\psi_j^*(-x,t)=\overline{c_{1j}}e^{\overline{\eta_j}}$, $\phi_j(x,t)=c_{2j}e^{-\eta_j}$ and $\phi_j^*(-x,t)=\overline{c_{2j}}e^{\overline{-\eta_j}}$, where $\eta_j=i\lambda_jx+2i\lambda_j^2t$ and $\overline{\eta_j}=i\overline{\lambda_j}x+2i\overline{\lambda_j}^2t$, $j=1,2,\cdots,N$, we can substitute them in the Nth iterated DT formula and obtain the N-soliton solution of NNLS equation.
\section{Dark and antidark solitons of NNLS equation}
To obtain dark soliton solutions of NNLS equation, we choose plane wave solution as the seed solution, that is $q(x,t)=a_1 e^{ibt}$ and $q^*(-x,t)=a_2e^{-ibt}$, $b=2a_1a_2$ to the defocusing NNLS equation, that is Eqs. (\ref{pct1}) and (\ref{pct2}) with $\sigma=-1$.  Substituting these solutions into the Lax pair equations (\ref{pctpsi1}) and solving them we obtain the basic solutions in the form 
\begin{align}
\psi_1(x,t)=&e^{ia_1a_2t}(c_1e^{\chi_1s_1}+c_2e^{-\chi_1s_1}),\nonumber\\
\phi_1(x,t)=&e^{-ia_1a_2t}(-\frac{c_1}{a_1}(\lambda_1+is_1)e^{\chi_1s_1}+\frac{c_2}{a_1}(-\lambda_1+is_1)e^{-\chi_1s_1}),\nonumber\\
\psi_1^*(-x,t)=&e^{ia_1a_2t}(\overline{c_1}e^{\chi_2s_2}+\overline{c_2}e^{-\chi_2s_2}),\nonumber\\
\phi_1^*(-x,t)=&e^{-ia_1a_2t}(-\frac{\overline{c_1}}{a_1}(\overline{\lambda_1}+is_2)e^{\chi_2s_2}+\frac{\overline{c_2}}{a_1}(-\overline{\lambda_1}+is_2)e^{-\chi_2s_2}),
\end{align}
where $\chi_1=x+2\lambda_1t$, $\chi_2=x+2\overline{\lambda_1}t$, $s_1=\sqrt{-\lambda_1^2+a_1a_2}$ and $s_2=\sqrt{-\overline{\lambda_1}^2+a_1a_2}$.  Substituting these basic solutions into the first iterated DT formula (\ref{p1q1new}) and simplifying the resultant expressions with $\gamma_1=\frac{c_1}{c_2}$ and $\gamma_2=\frac{\overline{c_1}}{\overline{c_2}}$, we arrive at
\begin{align}
q[1](x,t)=&a_1e^{2ia_1a_2t}\left(1+2(\overline{\lambda_1}-\lambda_1)\frac{A}{D_2}\right),\nonumber\\
q[1]^*(-x,t)=&a_2e^{-2ia_1a_2t}\left(1-\frac{2}{a_1a_2}(\overline{\lambda_1}-\lambda_1)\frac{B}{D_2}\right),
\label{onedark1}
\end{align}
where 
\begin{align}
A=&\gamma_1\gamma_2e^{s_1\chi_1+s_2\chi_2}+\gamma_2e^{-s_1\chi_1+s_2\chi_2}+\gamma_1e^{s_1\chi_1-s_2\chi_2}+e^{-s_1\chi_1-s_2\chi_2},\nonumber\\
B=&\gamma_1\gamma_2p_1p_2e^{s_1\chi_1+s_2\chi_2}+\gamma_2p_3p_2e^{-s_1\chi_1+s_2\chi_2}+\gamma_1p_1p_4e^{s_1\chi_1-s_2\chi_2}+p_3p_4e^{-s_1\chi_1-s_2\chi_2},\nonumber\\
D_2=&\gamma_1\gamma_2(p_1-p_2)e^{s_1\chi_1+s_2\chi_2}+\gamma_2(p_3-p_2)e^{-s_1\chi_1+s_2\chi_2}-\gamma_1(p_4-p_1)e^{s_1\chi_1-s_2\chi_2}\nonumber\\&-(p_4-p_3)e^{-s_1\chi_1-s_2\chi_2},
\label{onedark2}
\end{align}
and $p_1=\lambda_1+is_1$, $p_2=\overline{\lambda_1}+is_2$, $p_3=\lambda_1-is_1$, $p_4=\overline{\lambda_1}-is_2$. 

\par According to the definition of dark soliton, the soliton solution with real functions asymptotically approaches constant value.  Therefore to identify dark soliton solution from (\ref{onedark1}) we must impose the conditions that $\lambda_1$ and $\overline{\lambda_1}$ are to be real and $a_1a_2>\lambda_1^2, \overline{\lambda_1}^2$. This solution (\ref{onedark1}) is more general than the one presented in \cite{dad}.  In contrast to bright soliton, the dark soliton solution of NNLS equation exhibits stable propagation for $q[1](x,t)$ and $q[1]^*(-x,t)$ independently.  The solution (\ref{onedark1}) consists of collision between two dark and/or antidark solitons.  Interestingly when $q[1](x,t)$ exhibits dark soliton $q[1]^*(-x,t)$ exhibits either antidark or dark soliton.  This contrasting behavior is yet to be seen in the literature.
\section{Collision dynamics in two dark and antidark soliton solution}
To analyze the collision dynamics of dark/antidark solitons we name the two solitons in $q[1](x,t)$ as $S_1^q$ and $S_2^q$ and the two solitons in $q[1]^*(-x,t)$ as $S_1^{q*}$ and $S_2^{q*}$.  Each soliton in $q[1](x,t)$ can take two forms, namely dark and antidark soliton form. In other words, one can get four different combinations of collision behaviours such as, (i) $S_1^q$: dark, $S_2^q$: dark, (ii) $S_1^q$: dark, $S_2^q$:  antidark, (iii) $S_1^q$: antidark, $S_2^q$: dark and (iv) $S_1^q$: antidark, $S_2^q$: antidark.  Similarly one can also get four different combinations of collisions between these solitons in $q[1]^*(-x,t)$, namely (i) $S_1^{q*}$: dark, $S_2^{q*}$: dark, (ii) $S_1^{q*}$: dark, $S_2^{q*}$:  antidark, (iii) $S_1^{q*}$: antidark, $S_2^{q*}$: dark and (iv) $S_1^{q*}$: antidark, $S_2^{q*}$: antidark.  Now combining $q[1](x,t)$ with $q[1]^*(-x,t)$ yields $2^2\times 2^2=16$ combinations of collisions in both $q[1](x,t)$ and $q[1]^*(-x,t)$.  Upon studying all the combinations we found that out of sixteen combinations only nine combinations are distinct and the remaining seven collisions mimic either one of these nine combinations of collisions.  For example, the two combinations in $q[1](x,t)$ such as $S_1^q$: dark, $S_2^q$: antidark soliton and $S_1^q$: antidark, $S_2^q$: dark soliton produce the same structure.  Hence it is sufficient to consider any one of the combinations among the two. We also plot $q[1]^*(-x,t)$ in the same manner.  Hence, in the following, we focus on only nine collision scenarios which produce different structures.\\
{\bf Collision Scenario 1:}\\
\begin{figure}[h!]
\includegraphics[width=\linewidth]{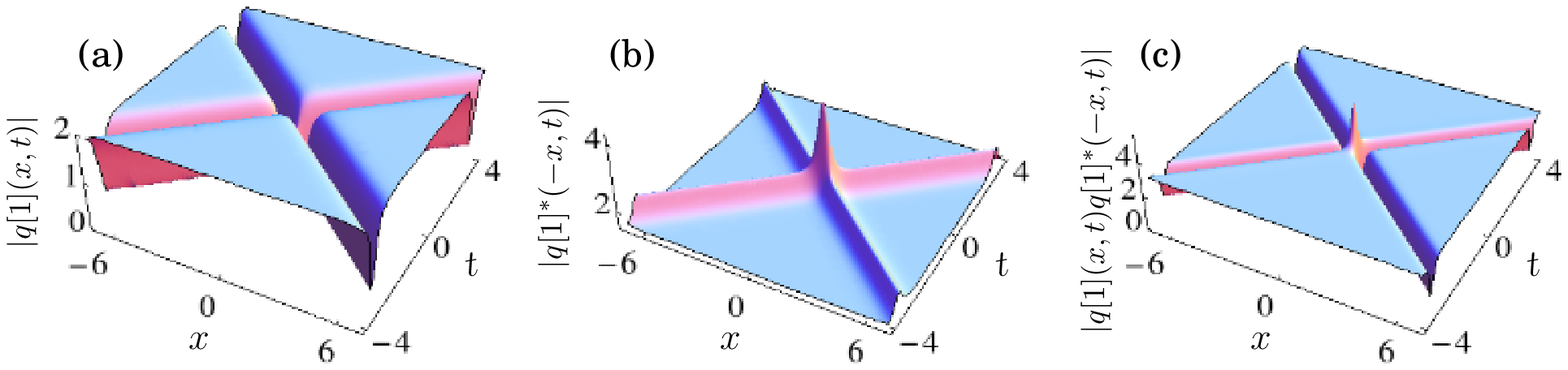}
\caption{(a) Two dark soltions collision of $|q[1](x,t)|$, (b) Two antidark solitons collision of $|q[1]^*(-x,t)|$, (c) Two dark solitons collision of $|q[1](x,t)q[1]^*(-x,t)|$ for the parameter values $\lambda_1=1$, $\overline{\lambda_1}=-1$, $a_1=2$, $a_2=1.5$, $\gamma_1=-1-.7i$, $\gamma_2=-0.5+.3i$.}
\end{figure}
Figure 3(a) shows collision between two dark solitons of $q[1](x,t)$ while $q[1]^*(-x,t)$ exhibits collision between two antidark solitons for the same parameter values which is shown in Fig. 3(b). Figure 3(c) shows the plot of $|q[1](x,t)q[1]^*(-x,t)|$, which gives two dark solitons collision.  Unlike local NLS case here the two dark soliton collision of $|q[1](x,t)q[1]^*(-x,t)|$ exhibits a hump during collison.\\ 
{\bf Collision Scenario 2:}\\
\begin{figure}[h!]
\includegraphics[width=\linewidth]{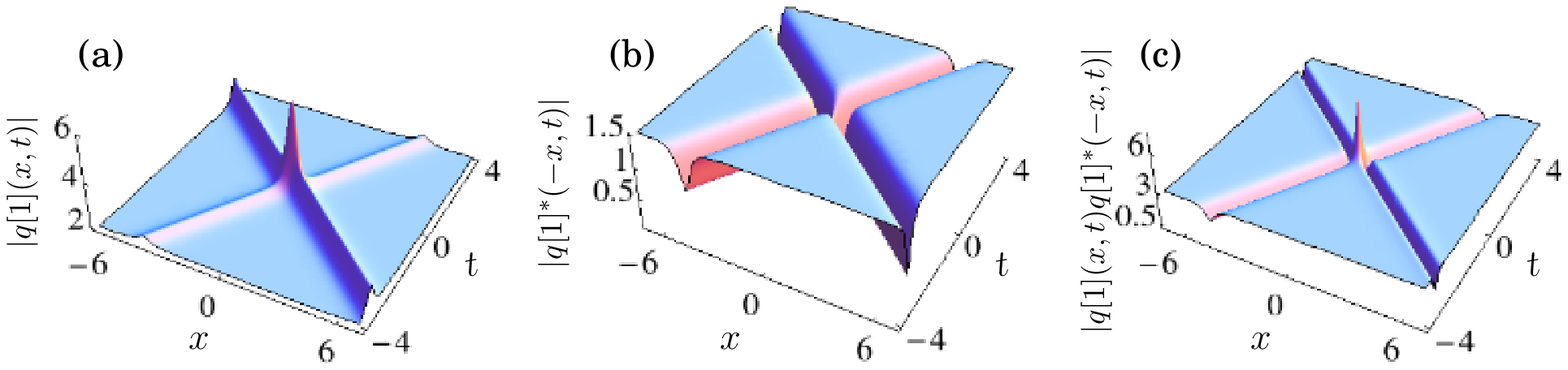}
\caption{ (a) Two antidark soltions collision of $|q[1](x,t)|$, (b) Two dark solitons collision of $|q[1]^*(-x,t)|$, (c) Two dark soltions collision of $|q[1](x,t)q[1]^*(-x,t)|$ for the parameter values $\lambda_1=1$, $\overline{\lambda_1}=-0.5$, $a_1=2$, $a_2=1.5$, $\gamma_1=1+i$, $\gamma_2=1.5-.3i$.}
\end{figure}
Figures 4(a) and 4(b) show contrasting structures compare to Figs. 3(a) and 3(b), that is $q[1](x,t)$ exhibits two antidark soliton collision and $q[1]^*(-x,t)$ gives two dark solitons collsion and Fig. 4(c) exhibits two dark solitons collision of $|q[1](x,t)q[1]^*(-x,t)|$ with a hump as in the previous case.\\
 {\bf Collision Scenario 3:}\\
\begin{figure}[h!]
\includegraphics[width=\linewidth]{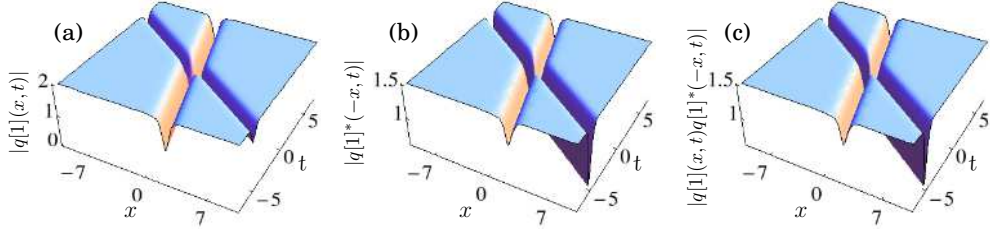}
\caption{(a) Two dark soltions collision of $|q[1](x,t)|$, (b) Two dark solitons collision of $|q[1]^*(-x,t)|$, (c) Two dark solitons collision of $|q[1](x,t)q[1]^*(-x,t)|$ for the parameter values $\lambda_1=0.2$, $\overline{\lambda_1}=1$, $a_1=2$, $a_2=1.5$, $\gamma_1=1+0.6i$, $\gamma_2=1-0.6i$.}
\end{figure}
Both $q[1](x,t)$ and $q[1]^*(-x,t)$ exhibit two dark solitons collision as shown in Figs. 5(a) and 5(b).  In this case also $|q[1](x,t)q[1]^*(-x,t)|$ produces two dark solitons collision without any hump as shown in Fig. 5(c). \\
{\bf Collision Scenario 4:}\\
\begin{figure}[h!]
\includegraphics[width=\linewidth]{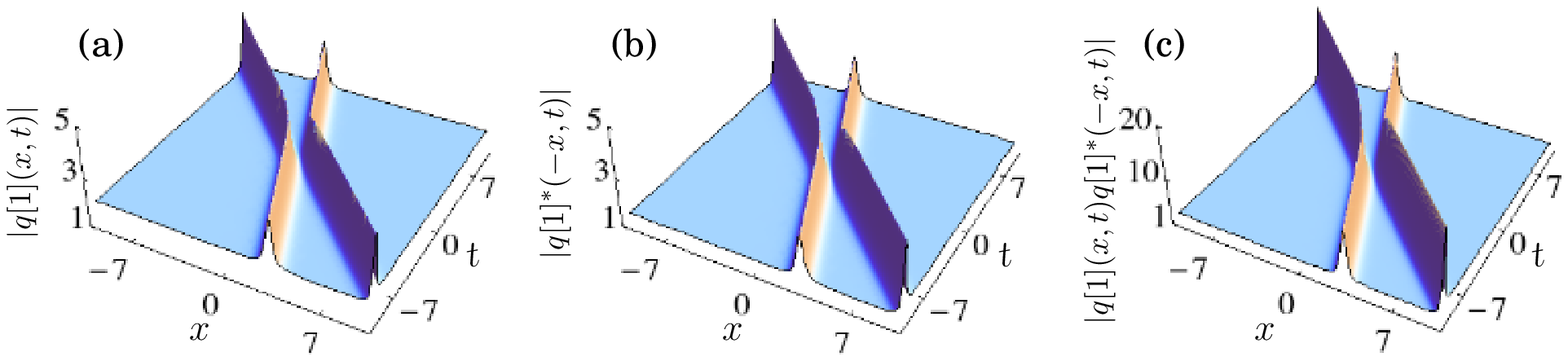}
\caption{(a) Two antidark soltions collision of $|q[1](x,t)|$, (b) Two antidark solitons collision of $|q[1]^*(-x,t)|$ and (c) Two antidark solitons collision of $|q[1](x,t)q[1]^*(-x,t)|$ for the parameter values $\lambda_1=0.5$, $\overline{\lambda_1}=0.2$, $a_1=2$, $a_2=1.5$, $\gamma_1=-1+0.6i$, $\gamma_2=1-0.6i$.}
\end{figure}
Both $q[1](x,t)$ and $q[1]^*(-x,t)$ exhibit collision between two antidark solitons shown in Figs. 6(a) and 6(b).  As we expected, $|q[1](x,t)q[1]^*(-x,t)|$ also exhibits two antidark solitons collision which is demonstrated in Fig. 6(c).\\
{\bf Collision Scenario 5:}\\
\begin{figure}[h!]
\includegraphics[width=\linewidth]{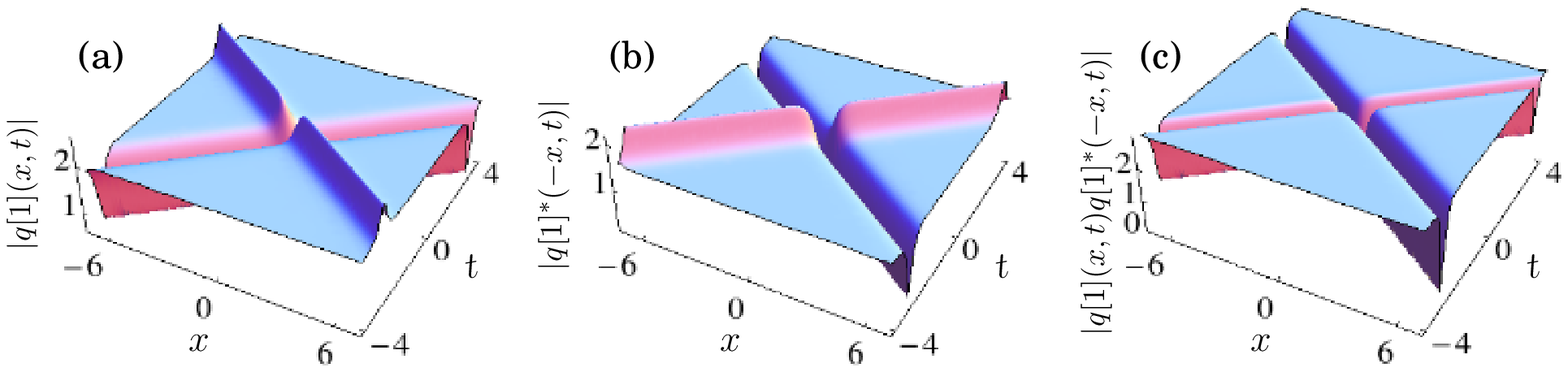}
\caption{(a) dark and antidark soltions collision of $|q[1](x,t)|$, (b) dark and antidark solitons collision of $|q[1]^*(-x,t)|$ and (c) Two dark solitons collision of $|q[1](x,t)q[1]^*(-x,t)|$ for the parameter values $\lambda_1=1.2$, $\overline{\lambda_1}=-1$, $a_1=2$, $a_2=1.5$, $\gamma_1=1+0.7i$, $\gamma_2=-0.5+0.3i$.}
\end{figure}
Figures 7(a) and 7(b) illustrate the collsions between dark and antidark solitons in both the components $q[1](x,t)$ and $q[1]^*(-x,t)$.  But when we plot $|q[1](x,t)q[1]^*(-x,t)|$, we get collision between two dark solitons, as shown in Fig. 7(c).\\
{\bf Collision Scenario 6:}\\
\begin{figure}[h!]
\includegraphics[width=\linewidth]{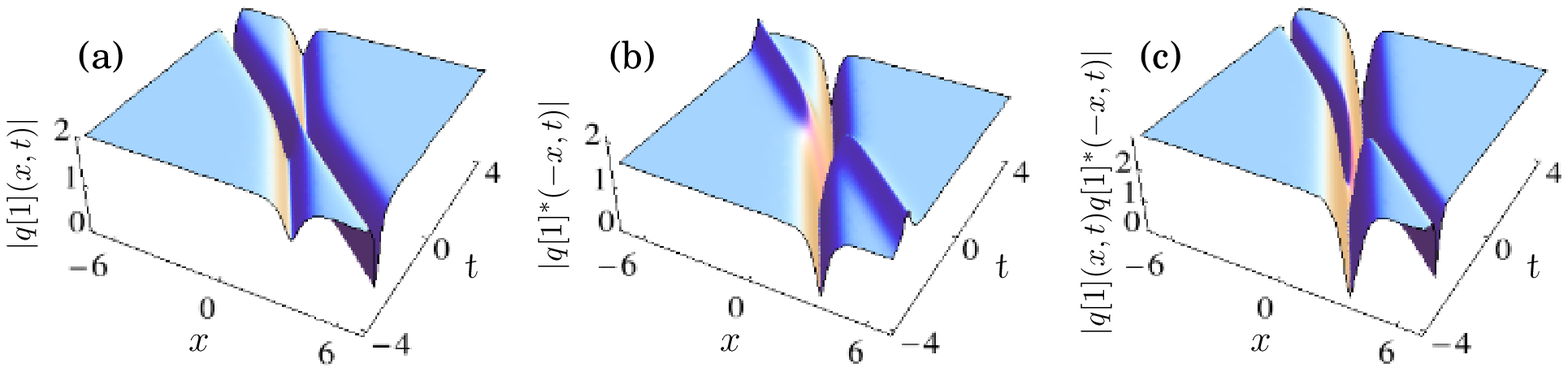}
\caption{(a) Two dark soltions collision of $|q[1](x,t)|$ and (b) dark and antidark solitons collision of $|q[1]^*(-x,t)|$ and (c) Two dark solitons collision of $|q[1](x,t)q[1]^*(-x,t)|$ for the parameter values $\lambda_1=1$, $\overline{\lambda_1}=0.5$, $a_1=2$, $a_2=1.5$, $\gamma_1=-1-0.7i$, $\gamma_2=-0.5+0.3i$.}
\end{figure} 
$q[1](x,t)$ exhibits two dark solitons collision and $q[1]^*(-x,t)$ gives collision between one dark and one antidark solitons which are shown in Figs. 8(a) and  8(b).  We then plot $|q[1](x,t)q[1]^*(-x,t)|$ for the same parameter values and obtain collision between two dark solitons as shown in Fig. 8(c). \\
{\bf Collision Scenario 7:}\\
\begin{figure}[h!]
\includegraphics[width=\linewidth]{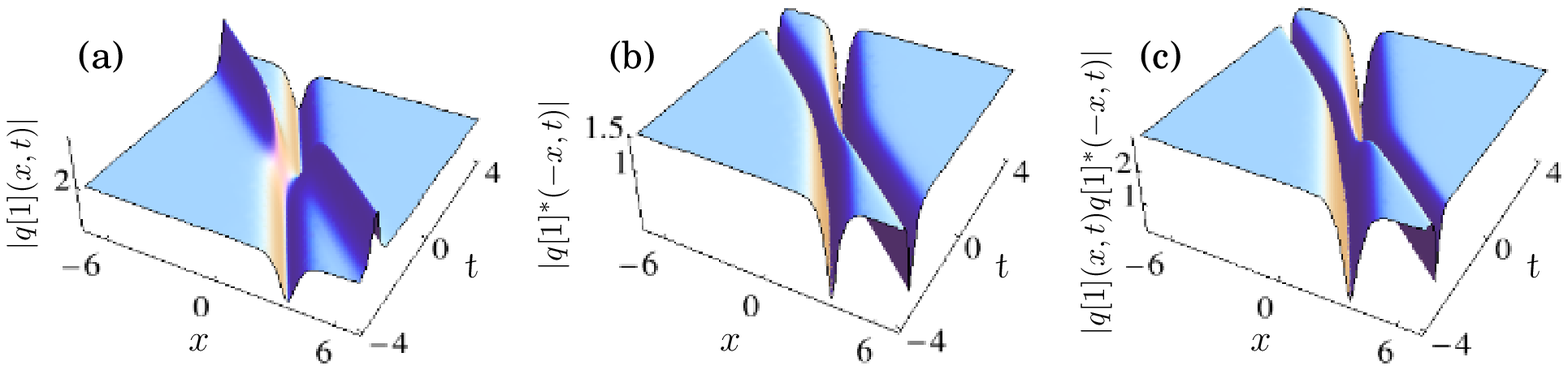}
\caption{(a) Dark and antidark soltions collision of $|q[1](x,t)|$, (b) two dark solitons collision of $|q[1]^*(-x,t)|$ and (c) two dark solitons collision of $|q[1](x,t)q[1]^*(-x,t)|$ for the parameter values $\lambda_1=1$, $\overline{\lambda_1}=0.5$, $a_1=2$, $a_2=1.5$, $\gamma_1=1+0.7i$, $\gamma_2=-0.5+0.3i$.}
\end{figure} 
In contrast to the previous collision scenario here we get dark and antidark soliton collision in $q[1](x,t)$ and two dark soliton collision in $q[1]^*(-x,t)$ which are illustrated in Figs. 9(a) and 9(b).  Here also we get two dark soliton collisions for $|q[1](x,t)q[1]^*(-x,t)|$ as demonstrated in Fig. 9(c).\\
{\bf Collision Scenario 8:}\\
\begin{figure}[h!]
\includegraphics[width=\linewidth]{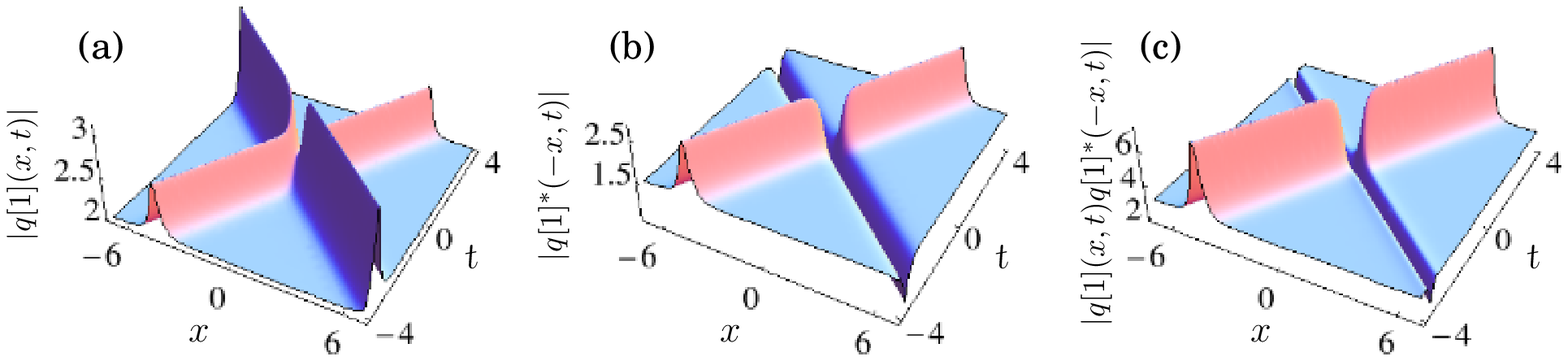}
\caption{(a) antidark-antidark soltion collision of $|q[1](x,t)|$, (b) dark-antidark soliton collision of $|q[1]^*(-x,t)|$ and (c) dark-antidark soliton collision of $|q[1](x,t)q[1]^*(-x,t)|$ for the parameter values $\lambda_1=1$, $\overline{\lambda_1}=-0.5$, $a_1=2$, $a_2=1.5$, $\gamma_1=1+i$, $\gamma_2=-1.5-0.3i$.}
\end{figure} 
For the parameter values given in Fig. 10 we get two antidark soliton collision in $q[1](x,t)$ and dark and antidark soliton collision in $q[1]^*(-x,t)$ which are illustrated in Figs. 10(a) and 10(b).  When we plot $|q[1](x,t)q[1]^*(-x,t)|$ for the same parameter values we get dark and antidark soliton collision that is displayed in Fig. 10(c). \\
{\bf Collision Scenario 9:}\\
\begin{figure}[h!]
\includegraphics[width=\linewidth]{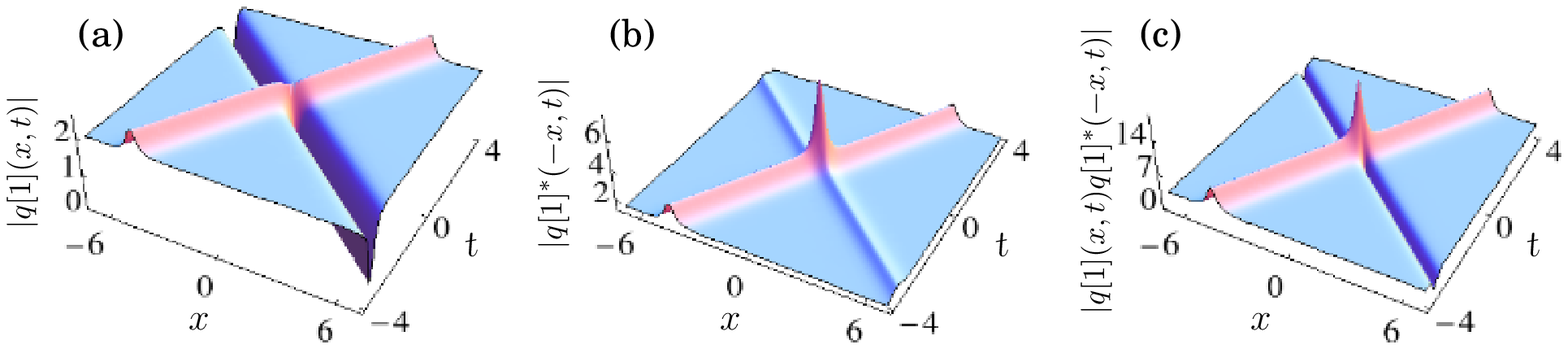}
\caption{(a) Dark and antidark soltions collision of $|q[1](x,t)|$, (b) two antidark solitons collision of $|q[1]^*(-x,t)|$ and (c) dark and antidark soliton collision of $|q[1](x,t)q[1]^*(-x,t)|$ for the parameter values $\lambda_1=1$, $\overline{\lambda_1}=-0.5$, $a_1=2$, $a_2=1.5$, $\gamma_1=-1-i$, $\gamma_2=-1.5-0.3i$.}
\end{figure} 
In contrast to the collision scenario 8, here $q[1](x,t)$ exhibits collision between dark and antidark solitons, see Fig. 11(a), and $q[1]^*(-x,t)$ gives two antidark soliton collision, see Fig. 11(b).  Plotting $|q[1](x,t)q[1]^*(-x,t)|$ gives dark and antidark collisions for the same parameter values which is given Fig. 11(c).
\par  To understand the above novel behaviours clearly we carry out appropriate asymptotic analysis in the following section.
\begin{table}[tp]%
\small
\begin{tabular}{|c|c|c|c|c|c|c|c|}
\hline
\multicolumn{2}{|c|}{$S_1^{q[1]^\pm}$} & \multicolumn{2}{|c|}{$S_2^{q[1]^\pm}$} & \multicolumn{2}{|c|}{$S_1^{q[1]^{*\pm}}$} & \multicolumn{2}{|c|}{$S_2^{q[1]^{*\pm}}$}\\
\hline
P.C & A.S & P.C & A.S & P.C & A.S & P.C & A.S \\
\hline
Im[$\gamma_1]>0$ & Dark &  Im[$\gamma_2]>0$ & Antidark & Im[$\gamma_1p_1/p_3]>0$ & Antidark & Im[$\gamma_2p_2/p_4]>0$ & Dark \\
\hline
Im[$\gamma_1]<0$ & Antidark &  Im[$\gamma_2]<0$ & Dark & Im[$\gamma_1p_1/p_3]<0$ & Dark & Im[$\gamma_2p_2/p_4]<0$ & Antidark \\
\hline
\end{tabular}
\caption{Asymptotic patterns of the solution (\ref{onedark1}) under different parametric conditions. (P.C - Parametric condition, A.S - Asymptotic soliton)}
\end{table}
\section{Asymptotic analysis for the defocussing case ($\sigma=-1$)}
The interpretation of the results in terms of actual motion of the soliton depends on the signs of the parameters $\lambda_1$ and $\overline{\lambda_1}$ which appear in the expressions $\chi_j$, $j=1,2$. To begin, let us assume that soliton 1 is in the vicinity of the line $x=-2\lambda_1t$. Now we change the frame co-moving with soliton 1 (coordinated by $\zeta$) by putting $x=\zeta-2\lambda_1t$.  As a result, for the soliton 2, we get $\chi_2=\zeta+2(\overline{\lambda_1}-\lambda_1)t$.  Considering the case $\overline{\lambda_1}>\lambda_1$ and $\overline{\lambda_1}>0$ we find that $\chi_2\to\pm\infty$ as $t\to\pm\infty$. Now we focus on soliton 2 which is located in the vicinity of the line $x=-2\overline{\lambda_1}t$.  We change the frame co-moving with soliton 2 (coordinated by $\zeta'$) as $x=\zeta'-2\overline{\lambda_1}t$. Then for the soliton 1, we get $\chi_1=\zeta'+2(\lambda_1-\overline{\lambda_1})t$.  Considering the case $\overline{\lambda_1}>\lambda_1$ and $\overline{\lambda_1}>0$, we find that $\chi_1\to\mp\infty$ as $t\to\pm\infty$. In the following, the superscripts in $q[1](x,t)$ and $q[1]^*(-x,t)$ represent the solitons (number(1,2)) and $\pm$ signs stand for $t\to\pm\infty$.\\\\
{\bf (i) Before Collision:}\\

Let us consider the limit $t\to-\infty$.  In the vicinity of $\chi_1\approx 0$, we have $\chi_2\to-\infty$, when $\overline{\lambda_1}>\lambda_1$, $\overline{\lambda_1}>0$.  Substituting these asymptotic values in the solution (\ref{onedark1}) we obtain the following results:\\\\
{\bf (a) Soliton 1:} 
\begin{align}
q[1](x,t)^{1-}=&a_1e^{2ia_1a_2t}\left(1+\frac{2(\overline{\lambda_1}-\lambda_1)\left(\gamma_1e^{s_1\chi_1}+e^{-s_1\chi_1}\right)}{-\gamma_1(-p_1+p_4)e^{s_1\chi_1}-(-p_3+p_4)e^{-s_1\chi_1}}\right),\nonumber\\
q[1]^*(-x,t)^{1-}=&a_2e^{-2ia_1a_2t}\left(1-\frac{\frac{2}{a_1a_2}(\overline{\lambda_1}-\lambda_1)\left(\gamma_1p_1p_4e^{s_1\chi_1}+p_3p_4e^{-s_1\chi_1}\right)}{-\gamma_1(-p_1+p_4)e^{s_1\chi_1}-(-p_3+p_4)e^{-s_1\chi_1}}\right),
\label{asym1}
\end{align}
and the squares of the absolute values are given by
\begin{align}
|q[1](x,t)^{1-}|^2=&a_1^2\left(1+\frac{2s_1\gamma_{1I}}{\gamma_{1R}\lambda_1-s_1\gamma_{1I}-\sqrt{a_1a_2}|\gamma_1|\cosh[2s_1\chi_1+\frac{\Delta^{1-}}{2}]}\right),\nonumber\\
|q[1]^*(-x,t)^{1-}|^2=&a_2^2\left(1+\frac{\frac{2s_1}{a_1a_2}(-2\gamma_{1R}\lambda_1s_1+\gamma_{1I}s_1^2-\gamma_{1I}\lambda_1^2)}{\gamma_{1R}\lambda_1-s_1\gamma_{1I}+\sqrt{a_1a_2}|\gamma_1|\cosh[2s_1\chi_1+\frac{\Delta^{1-}}{2}]}\right).\
\label{asym11}
\end{align}
From (\ref{asym11}) one can find the amplitudes 
\begin{align}
A^{1-}|_{q[1](x,t)}=&\frac{2a_1^2s_1\gamma_{1I}}{\gamma_{1R}\lambda_1-s_1\gamma_{1I}-\sqrt{a_1a_2}|\gamma_1|},\nonumber\\
A^{1-}|_{q[1]^*(-x,t)}=&\frac{\frac{2s_1a_2}{a_1}(-2\gamma_{1R}\lambda_1s_1+\gamma_{1I}s_1^2-\gamma_{1I}\lambda_1^2)}{\gamma_{1R}\lambda_1-s_1\gamma_{1I}+\sqrt{a_1a_2}|\gamma_1|},
\label{amp1}
\end{align}
and the phase as 
\begin{align}
\frac{\Delta^{1-}}{2}=\frac{1}{2}\ln\left(\frac{|\gamma_1|^2((s_1+s_2)^2+(\lambda_1-\overline{\lambda_1})^2)}{(s_1-s_2)^2+(\lambda_1-\overline{\lambda_1})^2}\right).
\label{phase1}
\end{align}
While plotting the solution (\ref{asym1}) we get dark soliton for $q[1](x,t)$ if $\text{Im}[\gamma_1]>0$ or antidark soliton for $q[1](x,t)$ if $\text{Im}[\gamma_1]<0$.  We obtain antidark soliton for $q^*[1](-x,t)$ if $\text{Im}[\gamma_1p_1/p_3]>0$ or dark soliton for $q[1]^*(-x,t)$ if $\text{Im}[\gamma_1p_1/p_3]<0$.  So we get two different conditions for $q[1](x,t)$ and $q[1]^*(-x,t)$ to obtain dark and antidark solitons. Due to this reason even if we have dark soliton for $q[1](x,t)$, $q[1]^*(-x,t)$ exhibits antidark soliton. From the amplitude and phase expressions (\ref{amp1}) and (\ref{phase1}) we can observe that $q[1](x,t)$ and $q[1]^*(-x,t)$ are evolving with different amplitudes but are having the same phase. \\\\
{\bf (b) Soliton 2:} \\
In the same limit $t\to-\infty$, in the vicinity of $\chi_2\approx 0$ we have $\chi_1\to\infty$, then we obtain
\begin{align}
q[1](x,t)^{2-}=&a_1e^{2ia_1a_2t}\left(1+\frac{2(\overline{\lambda_1}-\lambda_1)\left(\gamma_2e^{s_2\chi_2}+e^{-s_2\chi_2}\right)}{\gamma_2(p_1-p_2)e^{s_2\chi_2}-(-p_1+p_4)e^{-s_2\chi_2}}\right),\nonumber\\
q[1]^*(-x,t)^{2-}=&a_2e^{-2ia_1a_2t}\left(1-\frac{\frac{2}{a_1a_2}(\overline{\lambda_1}-\lambda_1)\left(\gamma_2p_1p_2e^{s_2\chi_2}+p_1p_4e^{-s_2\chi_2}\right)}{\gamma_2(p_1-p_2)e^{s_2\chi_2}-(-p_1+p_4)e^{-s_2\chi_2}}\right).
\label{asym2}
\end{align}
and the squares of their absolute values can be written as
\begin{align}
|q[1](x,t)^{2-}|^2=&a_1^2\left(1+\frac{2s_2\gamma_{2I}}{\gamma_{2R}\lambda_2-s_2\gamma_{2I}-\sqrt{a_1a_2}|\gamma_2|\cosh[2s_2\chi_2+\frac{\Delta^{2-}}{2}]}\right),\nonumber\\
|q[1]^*(-x,t)^{1-}|^2=&a_2^2\left(1+\frac{\frac{2s_2}{a_1a_2}(-2\gamma_{2R}\overline{\lambda_1}s_2+\gamma_{2I}s_2^2-\gamma_{2I}\lambda_2^2)}{\gamma_{2R}\overline{\lambda_1}-s_2\gamma_{2I}-\sqrt{a_1a_2}|\gamma_2|\cosh[2s_2\chi_2+\frac{\Delta^{2-}}{2}]}\right).\
\label{asym21}
\end{align}
From (\ref{asym21}), the amplitudes can be written as
\begin{align}
A^{2-}|_{q[1](x,t)}=&\frac{2a_1^2s_2\gamma_{2I}}{\gamma_{2R}\lambda_2-s_2\gamma_{2I}-\sqrt{a_1a_2}|\gamma_2|},\nonumber\\
A^{2-}|_{q[1]^*(-x,t)}=&\frac{\frac{2s_2a_2}{a_1}(-2\gamma_{2R}\overline{\lambda_1}s_2+\gamma_{2I}s_2^2-\gamma_{2I}\lambda_2^2)}{\gamma_{2R}\overline{\lambda_1}-s_2\gamma_{2I}-\sqrt{a_1a_2}|\gamma_2|},
\label{amp2}
\end{align}
and the phase is given by 
\begin{align}
\frac{\Delta^{2-}}{2}=\frac{1}{2}\ln\left(\frac{|\gamma_2|^2((s_1-s_2)^2+(\lambda_1-\overline{\lambda_1})^2)}{(s_1+s_2)^2+(\lambda_1-\overline{\lambda_1})^2}\right).
\label{phase2}
\end{align}
\par Now when we plot the solution (\ref{asym2}), we obtain antidark soliton for $q[1](x,t)$ if $\text{Im}[\gamma_2]>0$, and dark soliton for $q[1](x,t)$ if $\text{Im}[\gamma_2]<0$.  We get dark soliton for $q^*[1](-x,t)$ if $\text{Im}[\gamma_1p_2/p_4]>0$ or antidark soliton for $q^*[1](-x,t)$ if $\text{Im}[\gamma_1p_2/p_4]<0$.  Hence soliton 2 also exhibits two different conditions for $q[1](x,t)$ and $q^*[1](-x,t)$ to get dark and antidark solitons.  Here also we get the contrasting behaviour as in the previous case.   
\\

{\bf (ii) After Collision}\\
Now we consider the other limit $t\to\infty$. In the vicinity of $\chi_1\approx 0$, we have $\chi_2\to\infty$ and we get the following results:\\\\
{\bf (a) Soliton 1:}\\ 
\begin{align}
q[1](x,t)^{1+}=&a_1e^{2ia_1a_2t}\left(1+\frac{2(\overline{\lambda_1}-\lambda_1)\left(\gamma_1e^{s_1\chi_1}+e^{-s_1\chi_1}\right)}{\gamma_1(p_1-p_2)e^{s_1\chi_1}+(p_3-p_2)e^{-s_1\chi_1}}\right),\nonumber\\
q[1]^*(-x,t)^{1+}=&a_2e^{-2ia_1a_2t}\left(1-\frac{\frac{2}{a_1a_2}(\overline{\lambda_1}-\lambda_1)\left(\gamma_1p_1p_2e^{s_1\chi_1}+p_3p_2e^{-s_1\chi_1}\right)}{\gamma_1(p_1-p_2)e^{s_1\chi_1}+(p_3-p_2)e^{-s_1\chi_1}}\right),
\label{asym3}
\end{align}
and the squares of their absolute values are given by
\begin{align}
|q[1](x,t)^{1+}|^2=&a_1^2\left(1+\frac{2s_1\gamma_{1I}}{\gamma_{1R}\lambda_1-s_1\gamma_{1I}+\sqrt{a_1a_2}|\gamma_1|\cosh[2s_1\chi_1+\frac{\Delta^{1+}}{2}]}\right),\nonumber\\
|q[1]^*(-x,t)^{1+}|^2=&a_2^2\left(1+\frac{\frac{2s_1}{a_1a_2}(-2\gamma_{1R}\lambda_1s_1+\gamma_{1I}s_1^2-\gamma_{1I}\lambda_1^2)}{\gamma_{1R}\lambda_1-s_1\gamma_{1I}+\sqrt{a_1a_2}|\gamma_1|\cosh[2s_1\chi_1+\frac{\Delta^{1+}}{2}]}\right).\
\label{asym31}
\end{align}
From (\ref{asym31}), the amplitudes can be written as
\begin{align}
A^{1+}|_{q[1](x,t)}=&\frac{2a_1^2s_1\gamma_{1I}}{\gamma_{1R}\lambda_1-s_1\gamma_{1I}+\sqrt{a_1a_2}|\gamma_1|},\nonumber\\
A^{1+}|_{q[1]^*(-x,t)}=&\frac{\frac{2s_1a_2}{a_1}(-2\gamma_{1R}\lambda_1s_1+\gamma_{1I}s_1^2-\gamma_{1I}\lambda_1^2)}{\gamma_{1R}\lambda_1-s_1\gamma_{1I}+\sqrt{a_1a_2}|\gamma_1|}.
\label{amp3}
\end{align}
and the phase is given by 
\begin{align}
\frac{\Delta^{1+}}{2}=\frac{1}{2}\ln\left(\frac{|\gamma_1|^2((s_1-s_2)^2+(\lambda_1-\overline{\lambda_1})^2)}{(s_1+s_2)^2+(\lambda_1-\overline{\lambda_1})^2}\right).
\label{phase3}
\end{align}
{\bf (b) Soliton 2:}\\ 
In the same limit $t\to\infty$, in the vicinity of $\chi_2\approx 0$ we have $\chi_1\to-\infty$, then we get
\begin{align}
q[1](x,t)^{2+}=&a_1e^{2ia_1a_2t}\left(1+\frac{2(\overline{\lambda_1}-\lambda_1)\left(\gamma_2e^{s_2\chi_2}+e^{-s_2\chi_2}\right)}{\gamma_2(p_3-p_2)e^{s_2\chi_2}-(-p_3+p_4)e^{-s_2\chi_2}}\right),\nonumber\\
q[1]^*(-x,t)^{2+}=&a_2e^{-2ia_1a_2t}\left(1-\frac{\frac{2}{a_1a_2}(\overline{\lambda_1}-\lambda_1)\left(\gamma_2p_3p_2e^{s_2\chi_2}+p_3p_4e^{-s_2\chi_2}\right)}{\gamma_2(p_3-p_2)e^{s_2\chi_2}-(-p_3+p_4)e^{-s_2\chi_2}}\right).
\label{asym4}
\end{align}
and the squares of their absolute values can be written as
\begin{align}
|q[1](x,t)^{2+}|^2=&a_1^2\left(1+\frac{2s_2\gamma_{2I}}{\gamma_{2R}\lambda_2-s_2\gamma_{2I}-\sqrt{a_1a_2}|\gamma_2|\cosh[2s_2\chi_2+\frac{\Delta^{2+}}{2}]}\right),\nonumber\\
|q[1]^*(-x,t)^{2+}|^2=&a_2^2\left(1+\frac{\frac{2s_2}{a_1a_2}(-2\gamma_{2R}\overline{\lambda_1}s_2+\gamma_{2I}s_2^2-\gamma_{2I}\lambda_2^2)}{\gamma_{2R}\overline{\lambda_1}-s_2\gamma_{2I}-\sqrt{a_1a_2}|\gamma_2|\cosh[2s_2\chi_2+\frac{\Delta^{2+}}{2}]}\right).\
\label{asym41}
\end{align}
From (\ref{asym41}) the amplitudes can be written as
\begin{align}
A^{2+}|_{q[1](x,t)}=&\frac{2a_1^2s_2\gamma_{2I}}{\gamma_{2R}\lambda_2-s_2\gamma_{2I}-\sqrt{a_1a_2}|\gamma_2|},\nonumber\\
A^{2+}|_{q[1]^*(-x,t)}=&\frac{\frac{2s_2a_2}{a_1}(-2\gamma_{2R}\overline{\lambda_1}s_2+\gamma_{2I}s_2^2-\gamma_{2I}\lambda_2^2)}{\gamma_{2R}\overline{\lambda_1}-s_2\gamma_{2I}-\sqrt{a_1a_2}|\gamma_2|},
\label{amp4}
\end{align}
and the phase is given by 
\begin{align}
\frac{\Delta^{2+}}{2}=\frac{1}{2}\ln\left(\frac{|\gamma_2|^2((s_1+s_2)^2+(\lambda_1-\overline{\lambda_1})^2)}{(s_1-s_2)^2+(\lambda_1-\overline{\lambda_1})^2}\right).
\label{phase4}
\end{align}
While plotting the solutions (\ref{asym3}) and (\ref{asym4}), we came across the same behaviour as in (\ref{asym1}) and (\ref{asym2}) respectively.  The parametric regions for dark and antidark solitons in both $q[1](x,t)$ and $q[1]^*(-x,t)$ are listed in Table 1.  From Eqs. (\ref{asym1}) - (\ref{asym4}), we can observe the following features. From the amplitude expressions, one can see the heights of the antidark solitons or the depths of the dark solitons are same before and after collisions, that is $A^{i-}|_{q[1](x,t)}=A^{i+}|_{q[1](x,t)}$ and $A^{i-}|_{q[1]^*(-x,t)}=A^{i+}|_{q[1]^*(-x,t)}$ .  The envelope velocity is also equal for $q[1]^{i-}$ and $q[1]^{i+}$ and also for $q[1]^{*i-}$ and $q[1]^{*i+}$, $i=1,2$, that is $v^{\pm}|_{q[1](x,t)}=2\lambda_1$ and $v^{\pm}|_{q[1]^*(-x,t)}=2\overline{\lambda_1}$.  From the Eqs. (\ref{phase1})-(\ref{phase4}), we observe phase shift of soliton 1 for both $q[1](x,t)$ and $q[1]^*(-x,t)$: $\frac{\Delta_1^+}{2}-\frac{\Delta_1^-}{2}=\ln\left(\frac{(s_1-s_2)^2+(\lambda_1-\overline{\lambda_1})^2}{(s_1+s_2)^2+(\lambda_1-\overline{\lambda_1})^2}\right)$ and phase shift of soliton 2 for both $q[1](x,t)$ and $q[1]^*(-x,t)$: $\frac{\Delta_2^+}{2}-\frac{\Delta_2^-}{2}=\ln\left(\frac{(s_1+s_2)^2+(\lambda_1-\overline{\lambda_1})^2}{(s_1-s_2)^2+(\lambda_1-\overline{\lambda_1})^2}\right)$.  These three features suggest that the solution (\ref{onedark1}) can describe the elastic two-soliton interactions on the continuous wave background, that is the interacting solitons can completely recover their shapes and velocities after interaction and experience only phase shifts for their envelops.  
\begin{figure}
\includegraphics[width=\linewidth]{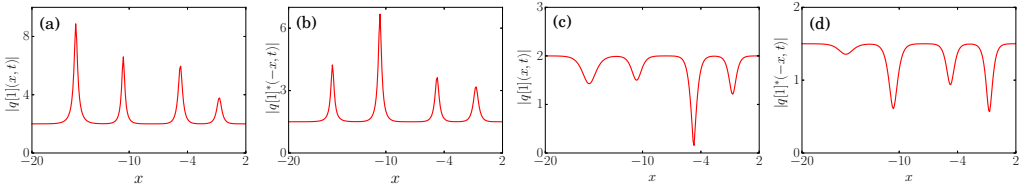}
\caption{(a) Structure of antidark four-soliton form of $|q[1](x,t)|$ at $t=5$, (b) Structure of antidark four-soliton form of $|q[1]^*(-x,t)|$ at $t=5$ for the parameter values $\lambda_1=0.2$, $\overline{\lambda_1}=1$, $\lambda_2=0.5$, $\overline{\lambda_2}=1.4$, $a_1=2$, $a_2=1.5$, $\gamma_1=1-0.6i$, $\gamma_2=1-0.6i$, $\gamma_3=0.2 + 0.5i$, $\gamma_4=-0.5 + 0.7i$ (c) Structure of dark four-soltion form of $|q[1](x,t)|$ at $t=5$, and (d) Structure of dark four-soltion form of $|q[1]^*(-x,t)|$ at $t=5$ for the parameter values $\lambda_1=0.2$, $\overline{\lambda_1}=1$, $\lambda_2=0.5$, $\overline{\lambda_2}=1.4$, $a_1=2$, $a_2=1.5$, $\gamma_1=-1+0.6i$, $\gamma_2=-1+0.6i$, $\gamma_3=-0.2 - 0.5i$, $\gamma_4=0.5 - 0.7i$}
\end{figure} 
\begin{figure}[h]
\includegraphics[width=\linewidth]{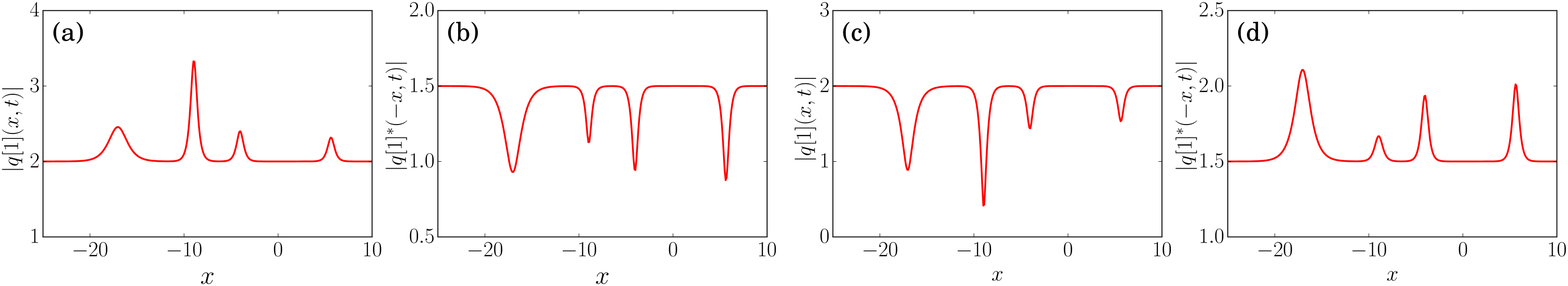}
\caption{(a) Structure of antidark four-soliton form of $|q[1](x,t)|$ at $t=5$, (b) Structure of dark four-soltion form of $|q[1]^*(-x,t)|$ at $t=5$ for the parameter values $\lambda_1=0.8$, $\overline{\lambda_1}=-0.5$, $\lambda_2=0.5$, $\overline{\lambda_2}=1.6$, $a_1=2$, $a_2=1.5$, $\gamma_1=-1-i$, $\gamma_2 = 1.5-0.3i$, $\gamma_3=0.2 + 0.05i$, $\gamma_4=0.5 + 0.7i$, (c) Structure of dark four-soltion form of $|q[1](x,t)|$ at $t=5$ and (d) Structure of antidark four-soliton form of $|q[1]^*(-x,t)|$ at $t=5$ for the parameter values $\lambda_1=0.8$, $\overline{\lambda_1}=-0.5$, $\lambda_2=0.5$, $\overline{\lambda_2}=1.6$, $a_1=2$, $a_2=1.5$, $\gamma_1=1+i$, $\gamma_2=-1.5 +0.3i$, $\gamma_3 = -0.2-0.05i$, $\gamma_4=-0.5 - 0.7i$}
\end{figure} 
\begin{figure}[h]
\includegraphics[width=\linewidth]{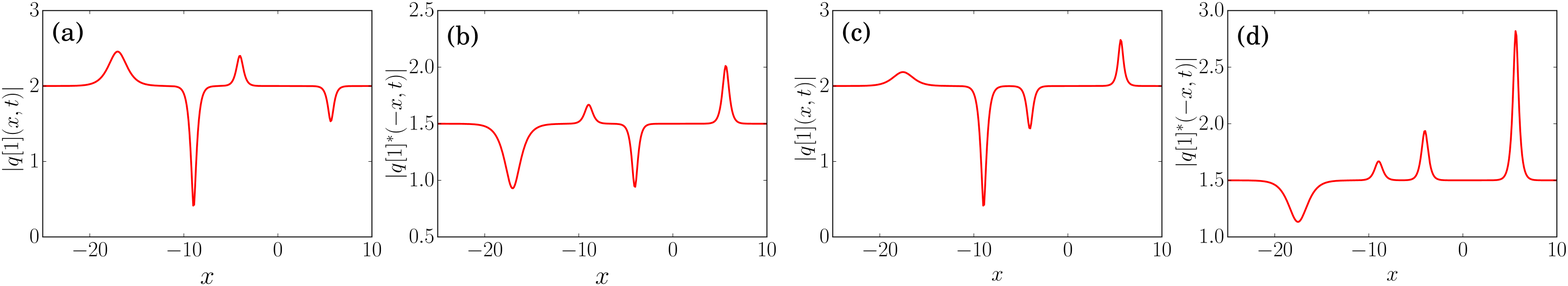}
\caption{(a) Structure of dark and antidark four-soliton form of $|q[1](x,t)|$ at $t=5$, (b) Structure of dark and antidark four-soliton form of $|q[1]^*(-x,t)|$ at $t=5$ for the parameter values $\lambda_1=0.8$, $\overline{\lambda_1}=-0.5$, $\lambda_2=0.5$, $\overline{\lambda_2}=1.6$, $a_1=2$, $a_2=1.5$, $\gamma_1=1+i$, $\gamma_2 = -1.5+0.3i$, $\gamma_3=0.2 + 0.05i$, $\gamma_4=0.5 + 0.7i$, (c) Structure of dark and antidark four-soliton form of $|q[1](x,t)|$ at $t=5$ and (d) Structure of dark and antidark four-soliton form of $|q[1]^*(-x,t)|$ at $t=5$ for the parameter values $\lambda_1=0.8$, $\overline{\lambda_1}=-0.5$, $\lambda_2=0.5$, $\overline{\lambda_2}=1.6$, $a_1=2$, $a_2=1.5$, $\gamma_1=1+i$, $\gamma_2=-1.5 -0.3i$, $\gamma_3 = -0.2-0.05i$, $\gamma_4=1.5 + 0.7i$}
\end{figure} 
\par To obtain symmetry preserving solution of two dark and antidark soliton solution one should consider $\lambda_1$ as complex variable such as $\lambda_1=\lambda_{1R}+i\lambda_{1I}$.  By choosing $a_1=a_2=\rho$, $\overline{\lambda}_1=\lambda_{1}^*$, $\gamma_2=\gamma_1^*$ and making $\lambda_{1R}=0$, the solution (\ref{onedark1}) becomes
\begin{eqnarray}
q[1](x,t)=\rho e^{2i\rho^2t}\left(1-\frac{2\lambda_{1I}(e^{2s_{r1}\chi_{r1}}+\gamma_1)(k_1e^{-2s_{r1}\omega_{r1}}+k_1^*\gamma_1^*)}{\rho^2e^{2s_{r1}(\chi_{r1}-\omega_{r1})}+\lambda_{1I}k_1\gamma_1e^{-2s_{r1}\omega_{r1}}+\lambda_{1I}k_1^*\gamma_1^*e^{2s_{r1}\chi_{r1}}+|\gamma_1|^2\rho^2}\right),\nonumber\\
\label{sym2}
\end{eqnarray}
where $s_{r1}=\sqrt{\rho^2-\lambda_{1I}^2}$, $\omega_{r1}=x-2\lambda_{1I}z$, $\chi_{r1}=x+2\lambda_{1I}z$, $k_1=\lambda_{1I}-is_{r1}$.  The solution (\ref{sym2}) is the symmetry preserving solution since $q[1]^*(-x,t)$ can be obtained by taking complex conjugate of (\ref{sym2}) and reversing the space variable in it.  This solution coincides with the solution reported in Ref. \cite{dad}.
\section{$2N$ dark and antidark soliton solution}
Finally, Let us consider $2N$ set of basic solutions such that
\begin{align}
\psi_j(x,t)=&e^{ia_1a_2t}(c_{1j}e^{\chi_{1j}s_{1j}}+c_{2j}e^{-\chi_{1j}s_{1j}}),\nonumber\\
\phi_j(x,)=&e^{-ia_1a_2t}(-\frac{c_{1j}}{a_1}(\lambda_j+is_{1j})e^{\chi_{1j}s_{1j}}+\frac{c_{2j}}{a_1}(-\lambda_j+is_{1j})e^{-\chi_{1j}s_{1j}}),\nonumber\\
\psi_j^*(-x,t)=&e^{ia_1a_2t}(\overline{c_{1j}}e^{\chi_{2j}s_{2j}}+\overline{c_{2j}}e^{-\chi_{2j}s_{2j}}),\nonumber\\
\phi_j^*(-x,t)=&e^{-ia_1a_2t}(-\frac{\overline{c_{1j}}}{a_1}(\overline{\lambda_j}+is_{2j})e^{\chi_{2j}s_{2j}}+\frac{\overline{c_{2j}}}{a_1}(-\overline{\lambda_j}+is_{2j})e^{-\chi_{2j}s_{2j}}),\;\;j=1,2,\cdots,N,
\end{align}
where $\chi_{1j}=x+2\lambda_jt$, $\chi_{2j}=x+2\overline{\lambda_j}t$, $s_{1j}=\sqrt{-\lambda_j^2+a_1a_2}$ and $s_{2j}=\sqrt{-\overline{\lambda_j}^2+a_1a_2}$. Substituting these basic solutions in the $N$th iterated DT formula (\ref{nit}), one can obtain $2N$ dark and antidark soliton solutions of NNLS equations (\ref{pct1}) and (\ref{pct2}).  For example, with the choice $N=2$ we can obtain four dark and antidark soliton solution of the NNLS equations.  Since the solution is too complicated, we present only the plots of four dark and antidark soliton solution.  Here also we visualize contrasting behaviors for $q[2](x,t)$ and $q[2]^*(-x,t)$.  For the illustration purpose, we plot some combinations of collisions in Figs. 12-14. In the four soliton solution case we can formulate $2^{4}\times 2^{4}$ different combinations of collision behaviours. By generalizing this to $2N$ soliton solution, we get $2^{2N}\times 2^{2N}$ different combinations of collisions.   
\section{Conclusions}
Using DT method we have constructed symmetry preserving and symmetry broken N-bright soliton solution for the NNLS equation of focussing type and given explicit one and two soliton solutions.  To construct these solutions, we have considered appropriate eigenfunctions in the DT method.  We have shown that due to the presence of PT-symmetric potential, the bright solitons exhibit unstable behaviour for the field and parity transformed complex conjugate field when we consider them separately.  However, they exhibit a stable behaviour when we combine the field and parity transformed complex conjugate field.  Further, we have obtained dark/antidark soliton solutions of NNLS equation of defocussing type and shown that they exhibit stable behaviour for the field and parity transformed complex conjugate field separately. Moreover, we have observed a contrasting behaviour between the envelope of the field and parity transformed complex conjugate envelope of the field. For a particular parametric choice we get dark (antidark) soliton for the field while the parity transformed complex conjugate field exhibits antidark (dark) soliton.  In other words, the complex conjugate envelope of the field evolves independently irrespective of the behaviour of the field.  By carrying out the relevant asymptotic analysis we have shown that both the field and complex conjugate envelope of the field exhibit dark/antidark solitons in different parametric regions.  The two dark/antidark soliton solution of both the field and PT transformed complex conjugate field possess sixteen combinations of collision scenario.  We have also constructed four dark/antidark soliton solution.  Since the solution is very lengthy we have demonstrated the nature of solutions only through the plots.  We have pointed out that four dark/antidark soliton solution possess $2^{4}\times 2^{4}$ combinations of collisions between the solitons.  Finally, we have indicated the structure of $2N$-dark/antidark soliton solution formula and pointed out that this solution may have $2^{2N}\times 2^{2N}$ combinations of collision.

\section*{Acknowledgements}
NVP wishes to thank the University Grants Commission (UGC), Government of India, for providing financial support through Dr.D.S. Kothari Post Doctoral Fellowship Scheme. The work of MS forms part of a research project sponsored by Science and Engineering Research Board (SERB), DST, Government of India under Grant No.EMR/2016/001818 and GR was supported by the JC Bose Fellowship and the UGC Advanced Centre for Mathematics. The work of ML is supported by a DST-SERB Distinguished Fellowship.

\end{document}